\documentclass[a4paper, amsfonts, amssymb, amsmath, reprint, showkeys, nofootinbib, twoside]{revtex4-1}
\def\apj{{ApJ}}                 % Astrophysical Journal
\def\apjl{{ApJ}}                % Astrophysical Journal, Letters
\def\prd{{Phys.~Rev.~D}}        % Physical Review D
\def\prl{{Phys.~Rev.~Lett.}}    % Physical Review Letters
\usepackage{amsmath,amstext}
\usepackage[T1]{fontenc}
\usepackage{amssymb}
\input{epsf}
\usepackage{graphicx}
\usepackage{ae,aecompl}

\usepackage{hyperref}
\usepackage{amsmath}
\usepackage{amssymb}
\usepackage{mathtools}
\usepackage{bm}
\usepackage{cleveref}
\usepackage{tensor}
\usepackage{braket}
\usepackage{enumitem}

\usepackage{color}

\newcommand{\spc}{\quad \quad \quad}

\def\be{\begin{equation}}
\def\ee{\end{equation}}
\def\beq{\begin{eqnarray}}
\def\eeq{\end{eqnarray}}

\begin{document}
\title{When the entropy has no maximum: A new perspective on the instability of the first-order theories of dissipation}
\author{L.~Gavassino, M.~Antonelli \& B.~Haskell}
\affiliation{Nicolaus Copernicus Astronomical Center, Polish Academy of Sciences, ul. Bartycka 18, 00-716 Warsaw, Poland}

\begin{abstract}
The first-order relativistic fluid theories of dissipation proposed by Eckart and Landau-Lifshitz have been proved to be unstable. They admit solutions which start in proximity of equilibrium and depart exponentially from it. We show that this behaviour is due to the fact that the total entropy of these fluids, restricted to the dynamically accessible states, has no upper bound. As a result, these systems have the tendency to constantly change according to the second law of thermodynamics and the unstable modes represent the directions of growth of the entropy in state space. We, then, verify that the conditions of stability of Israel and Stewart's theory are exactly the requirements for the entropy to have an absolute maximum. 
Hence, we explain how the instability of the first-order theories is a direct consequence of the truncation of the entropy current at the first order, which turns the maximum into a saddle point of the total entropy. Finally, we show that recently proposed first-order stable theories, constructed using more general frames, do not solve the instability problem by providing a maximum for the entropy, but, rather, are made stable by allowing for small violations of the second law.
\end{abstract}

\maketitle

\section{Introduction}

In the gravitational wave era \citep{AbbottNS1} it is necessary, more than ever, to have at our disposal relativistic hydrodynamic theories of dissipation which are well-suited for numerical implementation. Heat conduction and viscosity, in particular bulk viscosity \citep{AlfordRezzolla}, are thought to play a major role in neutron star mergers, and understanding these transport processes is necessary for a reliable interpretation of the data \citep{RadiceEjecta, RadiceEjecta2}. Special relativistic fluid dynamics is also an essential tool in high-energy nuclear physics, where it is successfully used to describe the quark-gluon plasma formed in heavy ion collisions \citep{Romatschke2017ejr}. 

The literature is rich of alternative theories of dissipation \citep{Israel_Stewart_1979,Liu1986,carter1991,Kovtun2019,BulkGavassino}, possessing different mathematical properties and physical insight, whose central purpose is mostly to solve the  pathological aspects of the minimal models of \cite{Eckart1940} and \cite{landau6}. However, born as the straightforward relativistic generalizations of Navier-Stokes and of the Fourier law \citep{Weinberg1971}, the theories of Eckart and Landau-Lifshitz still appear, intuitively, as a natural way of embedding viscosity and heat conduction in a relativistic framework. 
%We want to revisit this intuition.

The aim of this paper is to provide a physical perspective on the fundamental origin of incompatibility of this kind of straightforward Navier-Stokes approach with relativistic thermodynamics. Furthermore, understanding the cause of the incompatibility will allow us to interpret the physical content of the most important modern theories of dissipation depending on how they solve this structural inconsistency.

Among all the troublesome properties that a Navier-Stokes-type theory exhibits (which include acausality and differential equations of a mixed hyperbolic-parabolic form), we will focus here on the instability of the equilibrium.  \citet{Hiscock_Insatibility_first_order} have shown that, if homogeneous perfect-fluid configurations are slightly perturbed, the disturbance can grow with no bound, producing runaway solutions. This phenomenon is in contrast with our understanding of dissipation as the process which leads thermodynamic systems to converge to the equilibrium state as time goes to infinity.

The physical interpretation of this instability has never been completely clarified. Some qualitative studies were carried out by \citet{noto_rel} for the case of the heat conduction, which led to the conclusion that these anomalous behaviours might arise from an improper redistribution of the inertia between the particle and the entropy current. However, the apparently natural ``regular'' solution that these arguments seemed to suggest \citep{CarterRegular} has been proven to lead to a fluid model which is in turn unstable \citep{Olson1990}.

The most well known successful alternative to the Eckart and Landau-Lifshitz theories is the Israel-Stewart second-order theory \citep{Israel_Stewart_1979}, which has been shown to be causal and stable (for linear perturbations from equilibrium), if appropriate choices of the parameters are adopted \citep{Hishcock1983}. 
Interestingly, \cite{Bemfica_2018_conformi,Bemfica_2019_nonlinear,Kovtun2019} recently proved  that, if different hydrodynamic frames from those considered by \cite{Eckart1940} and \cite{landau6} are considered, stability and causality may be actually restored in a first-order theory. This unexpected result reveals that the pathological behaviour of the models of Eckart and Landau-Lifshitz does not arise directly from the first-order truncation, but must have a more subtle origin. 

Since both the second-order theories and the general-frame first-order theories admit the original formulations of Eckart and Landau-Lifshitz as particular cases, they can be stable only within a particular range of values of their parameters. 
In both cases the stability conditions can be obtained only though a detailed perturbative analysis about equilibrium \citep{Hishcock1983,Hoult2020} and the conditions one obtains lack an intuitive physical interpretation \citep{Poovuttikul2019}. This has lead some authors to consider the Israel-Stewart theory too complicated and artificial and to claim that the thermodynamic background is not sufficiently understood \citep{Van2012}. 

In this paper we clarify both the physical origin of the instability of the first-order theories and the thermodynamic meaning of the stability conditions of the models of \citet{Israel_Stewart_1979} and \citet{Bemfica_2019_nonlinear}. 

Throughout the paper we adopt the spacetime signature $ ( - , +, + , + ) $ and work in natural units $c=k_B=1$.

\section{Mathematical preliminaries}

To understand the origin of the instability of Eckart and Landau-Lifshitz theories we first need to know, from a mathematical perspective, why this phenomenon is not expected to occur in real dissipative systems. In this section we briefly recall the foundations of irreversible thermodynamics and set the stage for our discussion.

\subsection{Entropy as a Lyapunov function}
\label{SonoLya}

Given a set of dynamical variables $z^j$, which obey first-order equations of motion of the form
\begin{equation}
\dot{z}^j = \mathcal{F}^j(z^k),
\end{equation}
we say that a smooth (i.e. continuous, differentiable and with continuous partial derivatives) function 
\begin{equation}
S=S(z^j)
\end{equation}
is a Lyapunov function of the system if
\begin{enumerate}[label=(\roman*)]
\item $S$ admits an absolute maximum, i.e. there is a state defined by the values $z_{\text{eq}}^j$ such that 
\begin{equation}
S(z^j) \leq S (z_{\text{eq}}^j)  \spc \forall z^j ;
\end{equation} 
\item the maximum value of $S$ is reached only in $(z^j = z_{\text{eq}}^j)$, so the point of absolute maximum is unique;
\item $S$ is a non-decreasing function of time:
\begin{equation}
\dfrac{dS}{dt}= \dfrac{\partial S}{\partial z^j} \mathcal{F}^j \geq 0.
\end{equation}
\end{enumerate}
When the system admits a function $S$ of this kind, then the state $(z_{\text{eq}}^j)$ is an equilibrium state of the system. In fact, if the system has an initial condition
\begin{equation}
z^j(0) = z_{\text{eq}}^j,
\end{equation}
then $S$ has its maximum possible value at $t=0$. Since $S$ cannot decrease and $(z_{\text{eq}}^j)$ is the only state in which $S$ is maximum, then we necessarily have
\begin{equation}
z^j(t) = z_{\text{eq}}^j  \spc \forall t \geq 0.
\end{equation}
The state of equilibrium can also be shown to be Lyapunov stable \citep{lasalle1961stability}, namely for any $\epsilon>0$ there is a $\kappa >0$ such that, if
\begin{equation}
\delta_{jk} (z^j(0)-z_{\text{eq}}^j)(z^k(0)-z_{\text{eq}}^k) \leq \kappa^2,
\end{equation}
where $\delta_{jk}$ is the Kronecker delta symbol, then
\begin{equation}
\delta_{jk} (z^j(t)-z_{\text{eq}}^j)(z^k(t)-z_{\text{eq}}^k) \leq \epsilon^2
\end{equation}
for any $t \geq 0$. Intuitively, this means that if the system starts ``close enough'' to the equilibrium state, then it will remain ``close enough'' forever. This is due to the fact that, to run away from equilibrium, the system should make $S$ decrease, which is forbidden. Therefore, a system which is Lyapunov stable does not admit runaway solutions from equilibrium, but only solutions which converge to it, or evolve around it moving on surfaces at constant $S$.

In the kinetic theory of ideal gases, for finite systems governed by Boltzmann's transport equation, the entropy (defined as minus Boltzmann's H-functional, see e.g. \cite{huang_book}) satisfies all the requirements to be a Lyapunov function \citep{cercignani_book} over the state-space with fixed constants of motion, namely the total energy, linear momentum, angular momentum and possibly particle numbers or (in the case of a gas of quasi-particles) ``the superfluid velocity'' \citep{huang_book,Termo}. 
For this reason, once the constants of motion are assigned in the initial conditions, there is a unique equilibrium state and this state is guaranteed to be Lyapunov stable. 

In a generic thermodynamic system, the existence, uniqueness and stability of the global thermodynamic equilibrium are ensured only if 
(for fixed values of the constants of motion, which will appear in the equation of state of the system \cite{GavassinoTermometri2020})
the entropy is a Lyapunov function,
% \footnote{
% Again, for fixed values of the constants of motion, which in the end will appear in the global equation of state of the system \citep{GavassinoTermometri2020,Termo}.
% }
see \cite{Prigogine1978} and \cite{Beretta1986}. 
For this to be true, all the defining conditions (i,ii,iii) need to be fulfilled. 
When hydrodynamic models are formulated, however, the validity of the third requirement (which is nothing but the second law of thermodynamics) is usually enforced by construction, while in many cases the conditions (i) and (ii) are not.  

The main goal of this paper is to show that the instability of the Eckart and Landau-Lifshitz first-order theories is the result of the fact that in these theories the entropy does not have a maximum value and it may diverge even if the total energy and momentum of the system are conserved. Therefore, the existence of runaway solutions has a clear thermodynamic origin and is rooted in the fact that it is favourable for the system to depart from the perfect fluid state because this leads to an increase of entropy.  

\subsection{The degrees of freedom of the theory}\label{genericoAed}

Since our aim is to study the entropy as a function over all the configurations that the system is allowed explore, it is necessary to analyse this configuration space in detail. Throughout the paper we will consider a fluid whose energy-momentum tensor can be decomposed into a perfect-fluid part and a  non-equilibrium deviation $\mathfrak{T}^{\mu \nu}$ as follows
\begin{equation}
T^{\mu \nu} = (\rho +P)u^\mu u^\nu +Pg^{\mu \nu} + \mathfrak{T}^{\mu \nu}.
\end{equation}
The fluid is supposed not to interact with any other external field, so that  we can impose energy-momentum conservation
\begin{equation}\label{burp}
\nabla_\mu T^{\mu \nu}=0 \, .
\end{equation}
With the only exception of subsections \ref{BIs}, \ref{BIS2} and \ref{sisisisis}, we will always  assume for simplicity that the particle number is not conserved (zero chemical potential) and that the fluid in thermodynamic equilibrium behaves as an ideal gas of ultra-relativistic particles. 
Therefore, the equilibrium pressure $P$ and the equilibrium internal energy $\rho$ are related by the kinetic identity
\begin{equation}\label{kin}
P = \dfrac{1}{3} \rho
\end{equation}
and a radiation-type equation of state
\begin{equation}\label{EOS}
\rho = a_R \Theta^4
\end{equation}
holds, where $\Theta$ is the (equilibrium) temperature and $a_R$ is a constant. 
The rest-frame (equilibrium) entropy density $s$ can be obtained from the Euler relation
\begin{equation}\label{UEU}
s\Theta = \rho +P,
\end{equation}
which immediately implies
\begin{equation}\label{sSs}
s = \dfrac{4}{3} a_R \Theta^3 \, .
\end{equation}
This choice of fluid is made just to have a reference model in which all the calculations can be easily performed analytically. In fact, our purpose is not to give another proof of the instability of the first-order theories (which is a well-known fact), but to understand its thermodynamic meaning. Our simplified model, therefore, will only serve as a guiding example to the mechanisms of the instability, but the most important results of the paper will be shown (when necessary) to hold in full generality.

The total flow of entropy is assumed to be described by an entropy four-current
\begin{equation}
s^\mu = s u^\mu + \sigma^\mu,
\end{equation}
where $\sigma^\mu$ is a non-equilibrium contribution, which in general vanishes when $\mathfrak{T}^{\mu \nu}=0$. 
The second law of thermodynamics has the local form \citep{degroot_book,Israel_1981_review}
\begin{equation}\label{bnigckfml}
\nabla_\mu s^\mu \geq 0.
\end{equation}
In the following, we will work for simplicity in a flat spacetime with global inertial coordinates. 
Therefore, assuming that the fluid occupies a finite volume, we define the total entropy of the system at a given time as
\begin{equation}\label{bucatini}
S = \int s^0 \, d_3 x.
\end{equation}
Equation \eqref{bnigckfml}, then, implies   
\begin{equation}\label{zecondalegge}
\dfrac{dS}{dt} \geq 0,
\end{equation}
which is the second law in its global form.

In a general hydrodynamic model, the state of the fluid at a given time can be completely assigned by determining the values of all the hydrodynamic fields in that particular instant of time. Therefore, the state-space of the system is the set of all the possible fluid configurations. In our example, to specify the configuration of the fluid completely, we need to know at least four independent equilibrium quantities, such as $\Theta$ and the three spatial components $u^j$ of the four-velocity (from now on we adopt the notation that the index $j$ runs over the spatial components only: $j=1,2,3$). Regarding the degrees of freedom introduced by the dissipative terms, one needs to make a more careful analysis.

Let us focus, for definiteness, on the case of bulk viscosity, given by a choice of the non-equilibrium contributions of the form
\begin{equation}\label{bulkstress}
\mathfrak{T}^{\mu \nu} = \Pi \, (g^{\mu \nu}+u^\mu u^\nu).
\end{equation}
The scalar $\Pi$ is the viscous stress. In Newtonian hydrodynamics, the value of $\Pi$ is usually determined from the Navier-Stokes assumption
\begin{equation}\label{PIi}
\Pi = -\zeta \, \partial_j u^j.
\end{equation}
This implies that, in Newtonian hydrodynamics, if we assign the value of $u^j$ everywhere (on the hypersurface at constant time), then the value of $\Pi$ is automatically determined (analogous arguments hold for the shear stress and the heat flux, assuming the Fourier law). Therefore, in Newtonian hydrodynamics, the presence of dissipation does not introduce new degrees of freedom. 

In relativistic hydrodynamics, however, equation \eqref{PIi} cannot hold in every reference frame, because it is not covariant. The most trivial relativistic generalization of \eqref{PIi} is provided by the first-order prescription
\begin{equation}\label{jukilo}
\Pi = -\zeta \, \partial_\nu u^\nu.
\end{equation}
The presence of a term $\partial_t u^0$ implies that the knowledge of the four-velocity along the hypersurface at constant time is not enough to constrain the value of $\Pi$. Instead, the foregoing equation can be inverted as follows,
\begin{equation}\label{sxcdfv}
\partial_t u^0 = -\partial_j u^j -\dfrac{\Pi}{\zeta},
\end{equation} 
which shows that in relativity $\Pi$ can be considered a new degree of freedom of the model \citep{Hiscock_Insatibility_first_order} and \eqref{sxcdfv} is the new equation of motion that closes the system\footnote{
It can be rigorously proven that \eqref{jukilo} produces a new degree of freedom by verifying that if we replace $\Pi$ by $-\zeta \nabla_\nu u^\nu$ in the energy conservation relation $\partial_\mu T^{\mu 0}=0$, the resulting equation has a second order term $\partial_t \partial_t u^0$ (which has no Newtonian analogue). This implies that we need to specify also $\partial_t u^0$ (or equivalently $\Pi$) in the initial conditions.
}. Therefore, in a relativistic model for bulk viscosity, we need 5 independent hydrodynamic fields to completely specify the state (i.e. $\Theta$, $u^j$ and $\Pi$).

This  relativistic enlargement of the state-space occurs whenever one constructs, in a preferred reference frame, a parabolic equation of the kind $(A\partial_t+B\partial_x^2)f=0$ and then moves to a generic frame through a Lorentz boost. The derivatives in space become, in the new reference frame, linear combinations of derivatives in both space and time, producing a term proportional to $B\partial_{t}^2 f$ and therefore increasing the number of degrees of freedom of the model. More details about this mechanism can be found in appendix \ref{AAA}, where a brief analysis for the case of the diffusion equation (whose instability in relativity is formally identical to the instability of Landau-Lifshitz, see \cite{Kost2000}) is provided. In the appendix we also give an intuitive explanation of the connection underlined by \cite{Hishcock1983} between stability and causality (although the problem of causality is not explicitly addressed in the present work).
 
In summary, the approach of extended irreversible thermodynamics \citep{Stewart_1977,Jou_Extended} of treating the dissipative terms as degrees of freedom is unavoidable in relativity. In this sense, there is no conceptual difference between the second order theories (where the dissipative terms are promoted to degrees of freedom explicitly) and the first order theories (where, for the case of Eckart and Landau-Lifshitz, this is hidden behind the fact that in the rest frame of the fluid element there are no derivatives in time). 

We will show that this inevitable extension of the state space is the origin of the instability. 
In fact, it produces a new class of available thermodynamic states of the total fluid (which have no Newtonian analogue), and this generates novel paths in the state space in which the entropy can grow without any bound.

\section{Instability of the Eckart theory of heat conduction}\label{EcKo}

The first example we examine is the model for heat conduction proposed by \citet{Eckart1940}. 

\subsection{The instability mechanism}\label{the_prrof}

An energy-momentum tensor of the form
\begin{equation}\label{TTT}
T^{\mu \nu} = (\rho +P)u^\mu u^\nu +P g^{\mu \nu} + q^\mu u^\nu + u^\mu q^\nu
\end{equation}
is assumed. The four-vector $q^\mu$ is the heat flow and satisfies the geometrical constraint
\begin{equation}\label{uq}
u_\mu q^\mu =0 \, ,
\end{equation}
%We also assume for simplicity that the particle number is not conserved (zero chemical potential) and that the fluid is an ideal gas of ultra-relativistic particles. Therefore the pressure $P$ and the internal energy $\rho$ are related by the kinetic identity
%\begin{equation}\label{kin}
%P = \dfrac{1}{3} \rho
%\end{equation}
%and a radiation-type equation of state
%\begin{equation}\label{EOS}
%\rho = a_R \Theta^4
%\end{equation}
%holds, where $\Theta$ is the temperature and $a_R$ is a constant. The rest-frame entropy density $s$ can be obtained from the Euler relation
%\begin{equation}\label{UEU}
%s\Theta = \rho +P,
%\end{equation}
%which immediately implies
%\begin{equation}\label{sSs}
%s = \dfrac{4}{3} a_R \Theta^3.
%\end{equation}
while the entropy four-current is postulated to be
\begin{equation}\label{entropppP}
s^\mu = su^\mu + \dfrac{1}{\Theta} q^\mu.
\end{equation}
Therefore, this theory is formulated in the general form presented in subsection \ref{genericoAed} with
\begin{equation}
\mathfrak{T}^{\mu \nu}= q^\mu u^\nu + u^\mu q^\nu  \spc \sigma^\mu = \dfrac{1}{\Theta} q^\mu.
\end{equation}
%Eckart's theory ensures by construction the energy-momentum conservation and the second law of thermodynamics in a local form:
%\begin{equation}\label{burp}
%\nabla_\mu T^{\mu \nu }=0  \spc \nabla_\mu s^\mu \geq 0.
%\end{equation} 
Let us consider a homogeneous portion of fluid. Then the equations \eqref{burp} and \eqref{bnigckfml} acquire the simpler form (we recall that we work in a flat spacetime with global inertial coordinates)
\begin{equation}\label{burp2}
\partial_t T^{0 \nu} =0  \spc  \partial_t s^0 \geq 0.
\end{equation} 
The first equation implies that the energy and momentum densities, defined respectively as
\begin{equation}\label{CCCooonnn}
\mathcal{E} := T^{00}  \spc  \mathcal{P}^j := T^{0j},
\end{equation}
are necessarily conserved during the evolution of homogeneous fluids. The second equation of \eqref{burp2} is the local version of \eqref{zecondalegge} for homogeneous systems. Note that in homogeneous fluids the quantity per unit volume which needs to increase to ensure the validity of the second law is not the rest-frame entropy density $s$, but the entropy density $s^0$, measured in the frame in which the fluid is homogeneous. It is important to keep this difference in mind, because we will see that the Lorentz contraction of volumes plays a role in the instability.
%increasing the total entropy arbitrarily if the fluid is able to accelerate while keeping the rest-frame entropy finite.
%In fact, given a 3D spacelike hyper-surface $\Sigma$, the total entropy contained in $\Sigma$ is given by the flux
%\begin{equation}\label{pappagallo}
%S = -\int_\Sigma s^\mu d\Sigma_\mu.
%\end{equation}
%This implies that if we take as $\Sigma$ a hyper-surface at constant time which crosses the fluid portion we find
%\begin{equation}
%S= \int s^0 d_3 x
%\end{equation}
%and equation \eqref{burp2}, then, automatically gives
%\begin{equation}
%\dfrac{dS}{dt} \geq 0.
%\end{equation}

Now our aim is to show that, for fixed values of $\mathcal{E}$ and $\mathcal{P}^j$, the density $s^0$ can become arbitrarily large by varying the remaining unconstrained hydrodynamic variables. In this way it will be automatically proven that the entropy has no upper bound and thus cannot be a Lyapunov function for the system.

It is sufficient to work with $\mathcal{P}^j =0$. Then, we can use the invariance under rotations of the fluid element to restrict ourselves to the case in which
\begin{equation}\label{vG}
u^0 = \gamma  \spc  u^1 =\gamma v  \spc u^2=u^3=0,
\end{equation}
where 
\begin{equation}
\gamma = \dfrac{1}{\sqrt{1-v^2}}.
\end{equation}
From \eqref{vG}, one can easily show that the constraint \eqref{uq} explicitly reads
\begin{equation}\label{q0}
q^0 = v \, q^1.
\end{equation}
Let us briefly count the degrees of freedom. According to the discussion of subsection \ref{genericoAed}, to specify the state of the fluid completely we need to know, for the case of heat conduction, the value of 5 variables ($v,\Theta,q^1,q^2,q^3$). However, as we said before, there are 4 constraints ($\mathcal{E},\mathcal{P}^1,\mathcal{P}^2,\mathcal{P}^3$). This means that the system is in principle allowed to visit a 1D manifold of states (which we can parametrize e.g. with $v$) compatibly with four-momentum conservation. So we need to study how $s^0$ varies along this manifold.

First of all, we note that the conditions $\mathcal{P}^2=\mathcal{P}^3=0$, combined with the third equation of \eqref{vG}, immediately imply
\begin{equation}
q^2 = q^3 =0.
\end{equation}
Now, recalling equation \eqref{TTT}, we need to employ the two remaining constraints
\begin{equation}\label{system}
\begin{split}
& \mathcal{E} = 4P \gamma^2 -P + 2\gamma q^0  \\
& \mathcal{P}^1 = 4P\gamma^2 v +\gamma q^1 + \gamma v \, q^0 =0 \\
\end{split}
\end{equation}
to write $\Theta$ and $q^1$ as functions of $v$. Note that we have used the kinetic identity \eqref{kin} to substitute $\rho$ with $3P$. With a little algebra, and with the aid of equation \eqref{q0}, one can show that the system \eqref{system} is equivalent to
\begin{equation}\label{system2}
 P = \dfrac{1+v^2}{3-v^2} \, \mathcal{E}  \spc q^1 = -\dfrac{4\gamma v}{3-v^2} \, \mathcal{E}. 
\end{equation}
The second equation can be used to rewrite the zeroth component of \eqref{entropppP} in the form
\begin{equation}\label{zaz}
s^0 = \dfrac{\gamma s}{1+v^2}.
\end{equation} 
Combining equations \eqref{kin}, \eqref{EOS} and \eqref{sSs} one can show that
\begin{equation}
s = 4 \bigg( \dfrac{a_R P^3}{3} \bigg)^{1/4}.
\end{equation}
Using this expression into \eqref{zaz}, together with the expression for the pressure given in \eqref{system2}, we finally obtain
\begin{equation}\label{gringo}
s^0 = \tilde{s} \, \bigg( 1-v^2 \bigg)^{-1/2} \bigg( 1+v^2 \bigg)^{-1/4} \bigg( 1- \dfrac{v^2}{3} \bigg)^{-3/4} ,
\end{equation}
where we have defined
\begin{equation}
\tilde{s}:= \dfrac{4}{3} \big(a_R \mathcal{E}^3 \big)^{1/4}.
\end{equation}
Now we immediately see that $s^0$ has no upper bound. In fact, as $v \rightarrow \pm 1$,
\begin{equation}
s^0 \longrightarrow +\infty.
\end{equation}
This proves that the entropy does not have a maximum, so it is not a Lyapunov function for Eckart's theory.

\subsection{Physical interpretation of the instability}

We can, now, provide a physical interpretation of the instability of Eckart's theory and why runaway solutions are admitted. Let us consider the homogeneous fluid configuration we presented in the previous subsection (with $\mathcal{P}^j=0$) and let us impose $v=0$. From \eqref{vG} and \eqref{system2} we find
\begin{equation}\label{equillll}
u^j =0  \spc P = \dfrac{1}{3} \mathcal{E}  \spc  q^j =0.
\end{equation}
These equations imply that the fluid is at rest ($u^j=0$), in local thermodynamic equilibrium ($q^j=0$), with rest-frame energy density $\rho = 3P = \mathcal{E}$. It is a well known result of kinetic theory \citep{cercignani_book} that this fluid configuration should correspond to the state of global thermodynamic equilibrium, i.e. of absolute maximum entropy, compatibly with the conservation of the total energy and momentum. This configuration should therefore be stable, because any physically allowed (i.e. compatible with the conservation laws) spontaneous deviation from it would result in a decrease of the total entropy. In the case of Eckart's theory, however, this is not the case.

\begin{figure}
\begin{center}
	\includegraphics[width=0.5\textwidth]{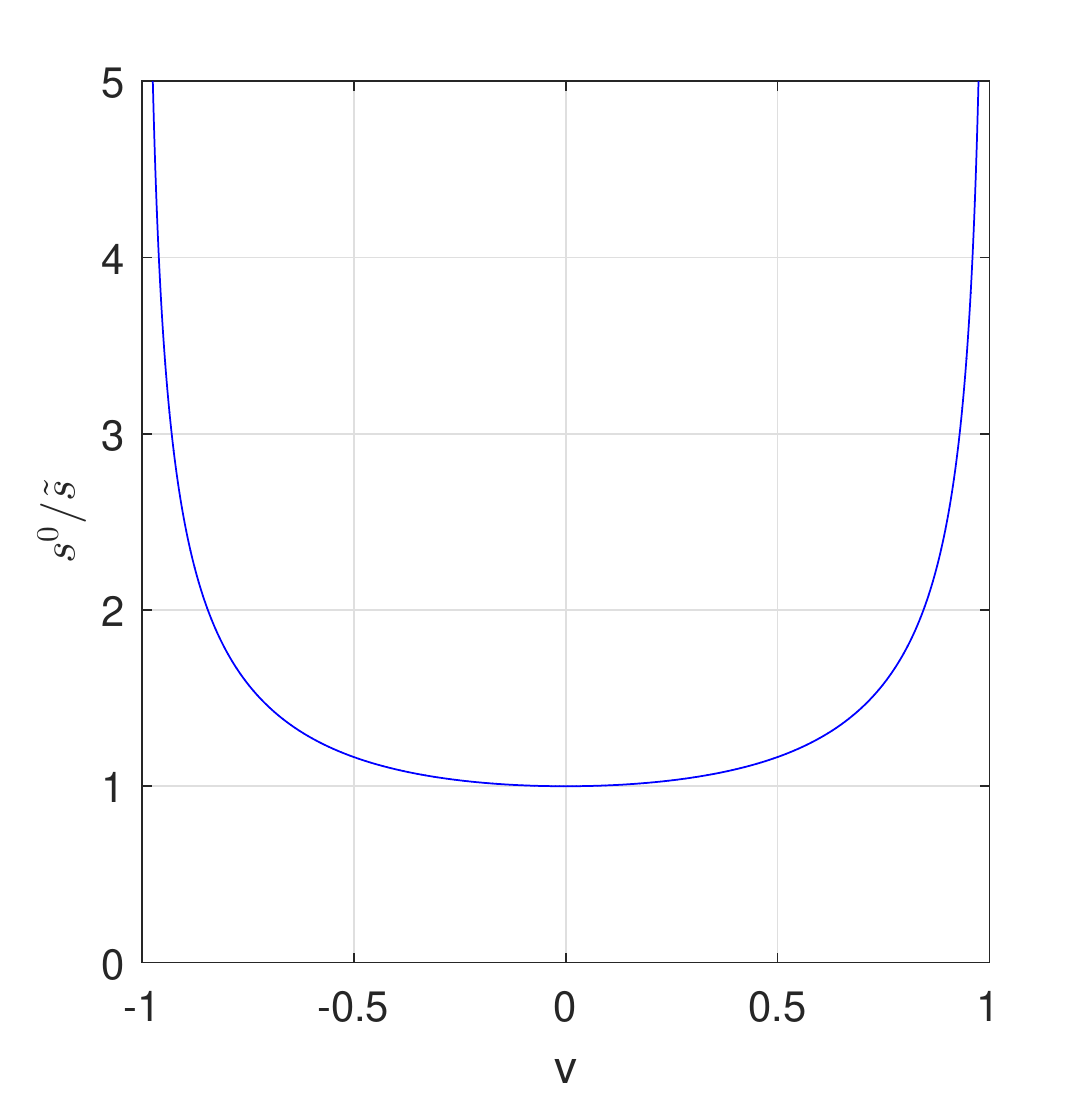}
	\caption{Plot of the Eckart normalised entropy per unit volume $s^0/\tilde{s}$ as a function of the velocity $v$, as given in equation \eqref{gringo}. As can be seen, the state  $v=0$ is not the maximum of the entropy density, but the minimum among the homogeneous configurations that the system is allowed to explore. 
	As discussed in the main text, this is the origin of the instability.
	}
	\label{fig:entropy}
	\end{center}
\end{figure}

To see this in more detail let us consider a state with a small $v>0$. From the second equation of \eqref{system2} we have
\begin{equation}\label{sbup}
q^1 \approx -\dfrac{4}{3} \mathcal{E} v.
\end{equation}
This corresponds to a configuration in which the fluid has accelerated in the positive direction $1$, using the heat flux $q^1$ as rocket fuel. In fact, the total momentum is still zero because the energy flow of the heat in the negative direction $1$ compensates the translational momentum of the fluid. We can compute the entropy in this state, expanding equation \eqref{gringo} to the second order, obtaining
\begin{equation}\label{S0}
s^0 \approx \tilde{s} \, \bigg( 1+ \dfrac{v^2}{2} \bigg).
\end{equation}
We see that the entropy is an increasing function of $v^2$. Now the origin of the instability is clear: since there is heat flux, entropy is necessarily produced, but if $s^0$ grows, then $v^2$ must increase, leading to a larger heat flux and therefore to a larger heat production. The fluid, then, accelerates more and more until it reaches the speed of light, where $s^0$ diverges, as can be seen in figure \ref{fig:entropy}. 

We have shown that the origin of the instability of Eckart's theory is fundamentally thermodynamic. Since the theory is constructed in a way to ensure the exact validity of the second law, the system will naturally evolve to the available state with maximum entropy. However, the perfect fluid state, identified by the condition $v=0$, is not the maximum of the entropy, but only a saddle point: restricting the entropy to the homogeneous configurations only, the perfect fluid state is the absolute minimum point. 

\subsection{The dynamics of the instability}

The problems of Eckart's theory are often attributed to the Fourier-type law \citep{Garcia2009}, namely the fact that the heat flow is expressed as \citep{Eckart1940}
\begin{equation}
\label{gigaq}
q^\mu = -k \Theta (g^{\mu \nu} + u^\mu u^\nu) \bigg[\dfrac{\nabla_\nu \Theta}{\Theta} + u^\sigma \nabla_\sigma u_\nu  \bigg],
\end{equation}
where $k$ is the conductivity coefficient.  
However, in our calculations we have never used explicitly the above expression. Since our analysis is purely thermodynamic (i.e. it deals only with the instantaneous properties of the system in a given state and not with its dynamical evolution in time), we see that the problem of Eckart's theory is more fundamental: it arises directly from the first-order expansion of the entropy current, equation \eqref{entropppP}. Equation \eqref{gigaq} produces the instability only because it encodes the second law in the model.

Nevertheless, we now show that the mechanism for the instability we have presented in the previous subsection produces the homogeneous unstable mode of Eckart's theory identified by \cite{Hiscock_Insatibility_first_order}, proving the complete consistency of our analysis with the first-order stability studies. 

Since we are working with global inertial coordinates, taking the component $\mu=1$ of equation \eqref{gigaq} gives
\begin{equation}
q^1 = -k \bigg( \partial_1 \Theta + u^1 u^\sigma \partial_\sigma \Theta + \Theta u^\sigma \partial_\sigma u^1  \bigg).
\end{equation}
We retain only the first order in $v$ and use the fact that the configuration is homogeneous to find
\begin{equation}\label{q1111}
q^1 = -k\Theta \dot{v}.
\end{equation}
Combining with \eqref{sbup} we get the exponential growth law
\begin{equation}\label{v+}
\dot{v} = \Gamma_+ v, 
\end{equation}
with a rate
\begin{equation}
\Gamma_+ = \dfrac{4\mathcal{E}}{3k\Theta}.
\end{equation}
This formula for $\Gamma_+$ is the formula of the imaginary frequency of the unstable mode identified by \citet{Hiscock_Insatibility_first_order}, cfr equation 52 therein. 
Now, we can use the existence of this runaway solution to prove that the entropy density, for small $v$, needs to have the second-order expansion given in equation \eqref{S0}. 

We consider the general equation for the entropy production of Eckart's theory, 
\begin{equation}\label{laproducoio}
 \nabla_\nu s^\nu = \dfrac{q^\nu q_\nu}{k\Theta^2}.
\end{equation} 
At the second order in $v$ it reduces to
\begin{equation}
\partial_t s^0 = \dfrac{q^1 q^1}{k\Theta^2},
\end{equation}
Using \eqref{q1111} and \eqref{v+} we can recast it into the form
\begin{equation}\label{partials0}
\partial_t s^0 = k\Gamma_+ v \dot{v}.
\end{equation}
However it is easy to verify that
\begin{equation}
\tilde{s} = k\Gamma_+,
\end{equation}
which, plugged into \eqref{partials0}, gives
\begin{equation}\label{perche}
\partial_t s^0 = \dfrac{\tilde{s}}{2} \, \partial_t (v^2).
\end{equation}
Then, considering that $s^0(v=0)=\tilde{s}$, we recover \eqref{S0}. Equation \eqref{perche} shows that the instability (i.e. the fact that $v^2$ grows) is a direct result of the second law (i.e. the fact that $s^0$ grows) and this can only be true if the entropy is not maximal in $v=0$. 

It is interesting to note that in this subsection we have followed an opposite path with respect to the previous ones. We have started directly from the first-order stability analysis of \cite{Hiscock_Insatibility_first_order} and we have studied the runaway solution directly. Using equation \eqref{laproducoio}, we have tracked how the entropy changes with time during the runaway, obtaining $s^0(t)$. Then, making the change of variable
\begin{equation}
s^0(t)=s^0(v(t)),
\end{equation}
we obtained the function $s^0(v)$ directly. 
Hence, we have shown that the unstable modes probe the convexity of the entropy near $v=0$. In this sense, the existence of the runaway solutions is the marker of its saddle point nature, showing that, since the perfect fluid state is not Lyapunov stable, the entropy cannot be a Lyapunov function of the system. 
This argument is fully general and applies beyond the toy-model we are considering here.

\subsection{Comparison with the Newtonian theory}

The fact that the runaway solutions are a purely relativistic effect is now evident. In fact, the dependence of $s^0$ on $v$ in the expansion \eqref{S0} is an order $v^2$ (where we recall that $v$ is measured in units of the speed of light). Furthermore, from \eqref{zaz} we also see that the convexity is due to the presence of coefficients like $\gamma$, which encodes the relativistic contraction of volumes, a phenomenon which does not exist in Newtonian physics. However, in the light of the discussion of subsection \ref{genericoAed}, it is  now possible to explain the fundamental mathematical difference between the two theories in more detail.

Let us consider the $0j$ component of the Eckart energy-momentum tensor \eqref{TTT}:
\begin{equation}
\mathcal{P}^j = (\rho +P)u^0 u^j+ q^0 u^j + q^j u^0.
\end{equation}
To obtain the Newtonian limit we need to take the limit
\begin{equation}
\rho \longrightarrow +\infty  \spc  u^0 \longrightarrow 1,
\end{equation}
obtaining the expression for the momentum density
\begin{equation}\label{PJJJ}
\mathcal{P}^j = \rho u^j.
\end{equation}
Hence, in the Newtonian limit 
\begin{equation}
\mathcal{P}^j =0  \quad \Longleftrightarrow \quad u^j=0.
\end{equation}
Since in  Newtonian physics the heat flux $q^\nu$ does not give any contribution to the momentum, this means that  the fluid configurations with $v\neq 0$ cannot be explored when the total momentum density is zero.  
Therefore, the mode that gives rise to the instability (which is the way in which the system can probe the profile of \eqref{gringo}) is not dynamically allowed and the fluid is forced to remain at rest. 

Again, this is just a reformulation of the statement that the state space of the relativistic fluid has a larger dimension with respect to the Newtonian one. A further confirmation comes from the fact that, as we see in \eqref{q1111}, the unstable mode is made possible only because of the presence of the time derivatives in the relativistic Fourier law \eqref{gigaq}, which makes $q^1$ a degree of freedom of the relativistic theory.

%in which the fluid is moving is one direction and the heat is flowing in the opposite direction, have necessarily non-zero momentum.
%
%This can also be seen directly from equation \eqref{gigaq}, which shows that the heat flux is the sum of two contributions,
%\begin{equation}
%q^\mu = q_F^\mu +q_I^\mu,
%\end{equation}
%where 
%\begin{equation}
%q_F^\mu = -k (g^{\mu \nu} + u^\mu u^\nu) \nabla_\nu \Theta
%\end{equation}
%is the analogue of the Fourier law (which survives in the Newtonian limit), and a second term
%\begin{equation}
%q_I^\mu = -k\Theta \, u^\sigma \nabla_\sigma u^\mu, 
%\end{equation}
%which vanishes in the Newtonian limit. The second term is a contribution which arises from the symmetry of the energy-momentum tensor, which in turn is the result of the equivalence between energy flows and momentum densities. It stems from the very fact that in relativity the heat itself contributes to the inertia of the fluid, invalidating \eqref{PJJJ}. We see from \eqref{q1111} that this term is responsible for the existence of the runaway solution, making it a purely relativistic effect.

\section{Instability of first-order bulk viscosity}

The instability mechanism we have presented in the previous section is not a specific feature of Eckart's model for heat conduction, but is a general problem of both \cite{Eckart1940} and \cite{landau6} theories. To see this, we perform an analogous study for the case of bulk viscosity. Note that both Eckart and Landau-Lifshitz theories treat this dissipative phenomenon in the same way, producing an instability which has been observed also in numerical simulations \cite{Etele2009}.

\subsection{The instability mechanism}

Let the stress-energy tensor be
\begin{equation}
T^{\mu \nu} = (\rho + P+\Pi)u^\mu u^\nu + (P+\Pi)g^{\mu \nu},
\end{equation}
and assume an entropy current
\begin{equation}
s^\mu = s u^\mu.
\end{equation}
The non-equilibrium stress correction $\mathfrak{T}^{\mu \nu}$ is, then, given by \eqref{bulkstress}, while the correction to the entropy current is $\sigma^\mu =0$. The reference perfect fluid is again assumed to be the one introduced in subsection \ref{genericoAed}.
We remark that the assumption that the fluid is an ultra-relativistic ideal gas must be considered in this section only a prescription for the equation of state and not a real microscopic interpretation, otherwise the bulk viscosity should vanish identically. Furthermore, again we focus on homogeneous configurations.

\begin{figure}
\begin{center}
	\includegraphics[width=0.5\textwidth]{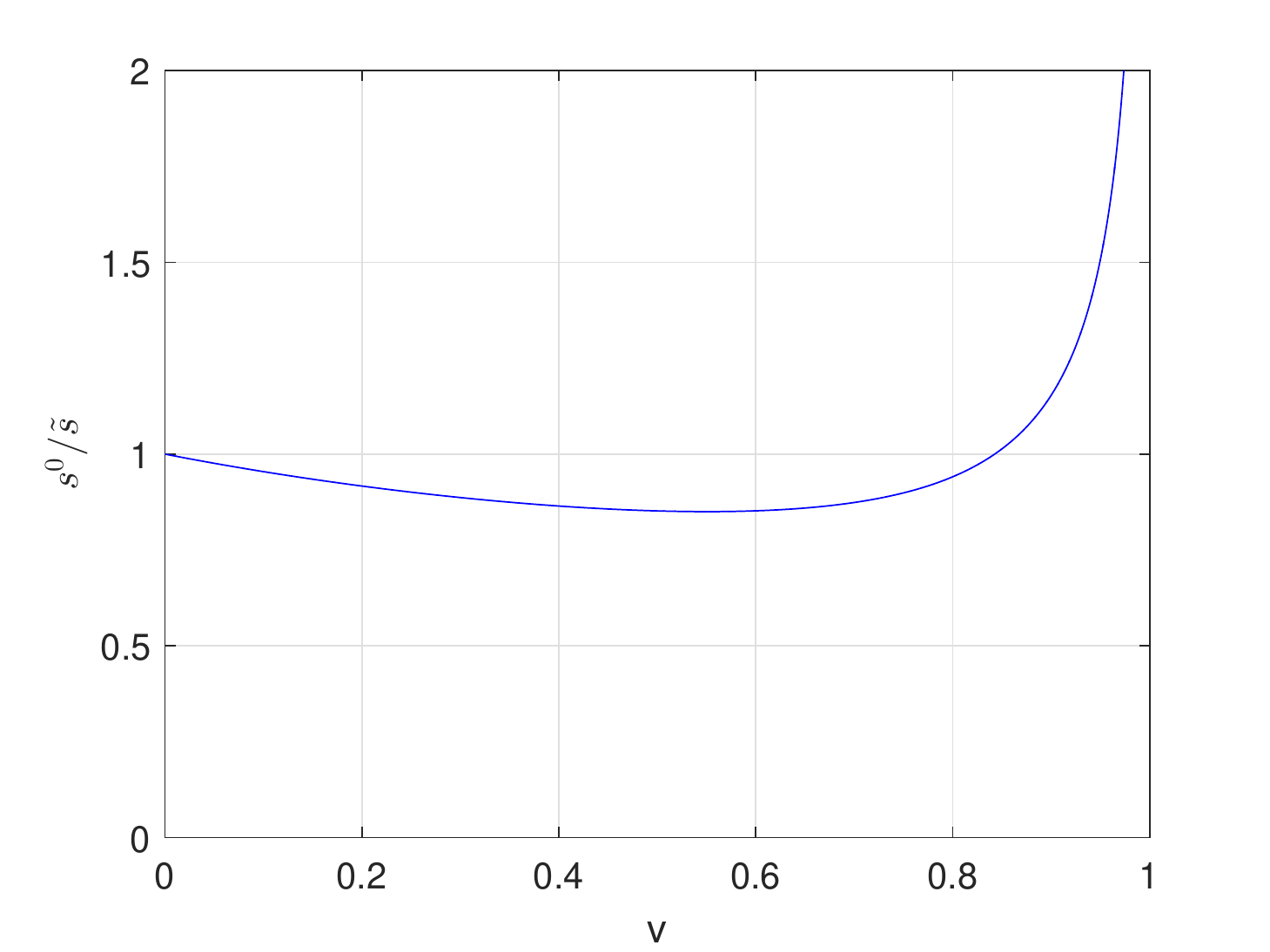}
	\caption{
	The normalised entropy per unit volume $s^0/\tilde{s}$ as a function of the velocity $v$, as given by equation \eqref{zfuf}, for $\mathcal{P}^1/\mathcal{E}=2/3$.  The state of local thermodynamic equilibrium is given by equation \eqref{equiaquil} (for $\mathcal{P}^1/\mathcal{E}=2/3$ we find $v_{eq} = 3-\sqrt{6} \approx 0.55 $) and corresponds to the minimum of $s^0/\tilde{s}$.
	The particular value $\mathcal{P}^1/\mathcal{E}=2/3$ has been selected for aesthetic reasons, the qualitative behaviour of the function does not depend on this choice.
	}
	\label{fig:bulk}
	\end{center}
\end{figure}

The line of reasoning is similar to that of subsection \ref{the_prrof}, with the difference that (as was shown by \citet{Hiscock_Insatibility_first_order}) the configuration in which the total momentum is zero is stable for homogeneous perturbations. Therefore, to see the unstable behaviour of the fluid we need to work with an unperturbed state in which the fluid is moving. Without any loss of generality, we can impose $\mathcal{E},\mathcal{P}^1 >0$, and
\begin{equation}
\mathcal{P}^2 = \mathcal{P}^3 =0.
\end{equation}
The degrees of freedom of the system now are $(v,\Theta,\Pi)$, while we have two relevant constraints $(\mathcal{E},\mathcal{P}^1)$. Thus, again, we are dealing with a 1D manifold of physically accessible states. 
The constraint equations read
\begin{equation}\label{syostum}
\begin{split}
& \mathcal{E} = (4P +\Pi) \gamma^2 -P -\Pi  \\
& \mathcal{P}^1 = (4P +\Pi)\gamma^2 v. \\
\end{split}
\end{equation}
These equations can be used to write $P$ and $\Pi$ as functions of $v$ along the curve, giving
\begin{equation}\label{e0orfoerfpokef}
P =\dfrac{1}{3} (\mathcal{E}-\mathcal{P}^1 v)  \spc  \Pi =\dfrac{\mathcal{P}^1}{v} \bigg( 1+\dfrac{v^2}{3} \bigg) -\dfrac{4}{3} \mathcal{E}. 
\end{equation} 
From the second equation we note that necessarily $v\neq 0$, which means that the system is dynamically allowed to exist only inside the open segment $0<v<1$. Setting $\Pi=0$ in the second equation we obtain the speed at which the fluid moves in local thermodynamic equilibrium,
\begin{equation}\label{equiaquil}
v_{\text{eq}} = \dfrac{2\mathcal{E} - \sqrt{4\mathcal{E}^2 - 3(\mathcal{P}^1)^2}}{\mathcal{P}^1}.
\end{equation}
The first equation of \eqref{e0orfoerfpokef} can be used to compute the entropy density $s^0$ as a function of $v$ (performing analogous calculations to those presented in the previous section) giving
\begin{equation}\label{zfuf}
s^0 = \tilde{s} \, \bigg( 1-v^2 \bigg)^{-1/2} \bigg( 1 - \dfrac{\mathcal{P}^1}{\mathcal{E}} v \bigg)^{3/4}.
\end{equation}
It is easy to show that $v_{\text{eq}}$ is the only stationary point of $s^0$. However, again, this point corresponds to the minimum of $s^0(v)$, as can be seen in figure \ref{fig:bulk}.

Now we are able to understand the instability of first-order bulk-viscous fluids. If the system starts with $v>v_{\text{eq}}$, then, since the entropy must grow, the fluid will be forced to accelerate until it reaches the speed of light. If, instead, it is prepared with $v<v_{\text{eq}}$, then it is forced to slow down until it stops. However, from the second equation of \eqref{e0orfoerfpokef}, we see that, as $v \longrightarrow 0$, the viscous stress diverges, $\Pi \longrightarrow \infty$, and, as $\Pi$ explodes, so does the entropy production. Therefore this system, of equations approaches in a finite time a singularity. 

It is clear why this instability mechanism is forbidden in Newtonian hydrodynamics: for these accelerations to be possible, the viscous stress needs to contribute to the total momentum, to ensure that the latter is constant during the evolution of the fluid.

\section{How to solve the problem}\label{HowToslove}

Since the theories of \citet{Israel_Stewart_1979} and \citet{Bemfica_2019_nonlinear} have been proved to be (conditionally) stable, they manage to avoid the instability mechanisms we presented in the previous sections. To understand how this is possible, we need first of all to clarify what does not work in the formulation of the theories of \citet{Eckart1940} and \citet{landau6} in the first place. In order to capture the physical essence of the problem, in this section we will deal with a thermodynamic toy-model which contains all the physical ingredients we need.

Consider a thermodynamic system with two degrees of freedom $(u,q)$, with no constants of motion and such that in a neighbourhood of the origin the entropy has the form (neglecting overall additive constants)
\begin{equation}\label{lkhypcs}
S = -u^2  +uq -q^2 \, .
\end{equation}
The function $S(u,q)$ is a quadratic form and can be rewritten in the diagonal representation
\begin{equation}
S = -\dfrac{1}{4} (u+q)^2 - \dfrac{3}{4}(u-q)^2 \, .
\end{equation} 
Therefore, $S$ admits a unique absolute maximum, given by
\begin{equation}
u=q=0,
\end{equation}
which is the stable equilibrium state of the system. 

Now,  let us assume  that  $|q| \ll |u|$ for every realistic initial condition of the system, 
so that we can expand $S$ to the first order in $q$ and neglect the contribution $q^2$ in \eqref{lkhypcs}. 
Hence, we obtain a first-order entropy
\begin{equation}
S_I = -u^2  +uq \, .
\end{equation}
The eigenvalues of the Hessian of $S_I $ are
\begin{equation}
\lambda_- =-1-\sqrt{2}  \spc  \lambda_+ = -1+\sqrt{2}.
\end{equation}
Since $\lambda_+ \lambda_- = -1<0$, the point $u=q=0$ is no longer the maximum of the entropy, but it is a saddle point. Therefore, we see that neglecting second-order contributions to the entropy one may destroy its nature of Lyapunov function. This is what happens when a first-order theory is constructed: the omission of second-order terms in the entropy current can lead to the removal of  contributions which are essential to determine the overall concavity of the total entropy, transforming its maximum into a saddle point.

There are two possible solutions to this problem. The first is to keep all the contributions to the second order. Clearly, if the microscopic input is realistic, all the terms should add up to give an entropy which is maximum in equilibrium (at least for small deviations from it), guaranteeing its Lyapunov stability. 
In fact, section \ref{IsrStew} is devoted to show that the conditions of stability of the second-order theory of \citet{Israel_Stewart_1979} are those which make the total entropy maximum in equilibrium. The same result holds also for the non-perturbative theories of \citet{carter1991}, \citet{lopez2011}, and \citet{BulkGavassino}, since they have the same stability properties as \cite{Israel_Stewart_1979}, see \cite{Priou1991}.

There is, however, an alternative approach. If one already knows that the equilibrium state must be $u=q=0$, then a first-order model for dissipation should just ensure that the system naturally evolves to this state and that 
\begin{equation}
\dfrac{dS_I}{dt} \geq 0
\end{equation}
only up to the first order in $q$. In other words, if $S_I$ is just the first order expansion of the ``real'' $S$, then its growth should be guaranteed except for terms of order $q^2$. In this way one has the freedom to construct the equations of motion for $u$ and $q$, tuning them in a way to ensure both the stability of the equilibrium, as an exact constraint, and the validity of the second law, as an approximate condition. Section \ref{Norio} is devoted to show that the possibility of small violations of the second law is the key to ensure the stability of the first-order theories in more general frames proposed by \citet{Bemfica_2018_conformi,Bemfica_2019_nonlinear} and \citet{Kovtun2019}.

We remark that in the present paper the terms \textit{first-order} and \textit{second-order theory} are interpreted according to the standard definition introduced by \citet{Hishcock1983} and currently used in textbooks \citep{rezzolla_book}. The terminology refers to the order of the expansion of the entropy four-current used to derive the equations of motion (a procedure which always requires one to move to the second order, even in the construction of first-order theories, as shown by \citet{Hishcock1983}), not to the order in the displacements from equilibrium of the final hydrodynamic equations. Therefore, according to this definition, Israel-Stewart remains a second-order theory even when it is linearised for small deviations about equilibrium (the equations retaining the Cattaneo-type structure, see appendix \ref{AAA}). As a consequence, the theories of heat conduction proposed by \citet{AnderssonLopezFirstOrder} and \citet{Van2012}, although referred to as first-order, are considered in the present paper as belonging to the class of second-order theories.

\section{Stability of the second order theories}\label{IsrStew}

We show that the conditions of stability of the theory of \citet{Israel_Stewart_1979} in the Eckart frame obtained by \citet{Hishcock1983} coincide with the condition for the total entropy of the fluid to be maximum in equilibrium (and thus to be a Lyapunov function).

\subsection{Brief summary of the theory}\label{BIs}

In the second-order theory one assumes that the stress-energy tensor of the fluid has the form
\begin{equation}
T^{\mu \nu} = (\rho + P+\Pi)u^\mu u^\nu + (P+\Pi)g^{\mu \nu} +q^\mu u^\nu +u^\mu q^\nu + \Pi^{\mu \nu}
\end{equation}
with
\begin{equation}
u_\mu q^\mu  =u_\mu \Pi^{\mu \nu}= \Pi^{[\mu \nu]}= \Pi\indices{^\mu _\mu}=0.
\end{equation}
The three dissipative contributions $q^\mu,\Pi,\Pi^{\mu \nu}$ are respectively heat flux, bulk viscosity and shear viscosity. Furthermore, it is assumed that there is a conserved particle current $n^\mu = n u^\mu$ (we work in the Eckart frame) such that
\begin{equation}\label{conservo}
\nabla_\mu n^\mu =0.
\end{equation}
The entropy current is expanded according to the logic of the extended irreversible thermodynamics approach \citep{Jou_Extended}:
\begin{equation}\label{secondoOrdine}
\begin{split}
 s^\mu = & su^\mu + \dfrac{q^\mu}{\Theta} -\dfrac{1}{2} \big(\beta_0 \Pi^2 +\beta_1 q^\nu q_\nu + \beta_2 \Pi^{\nu \rho} \Pi_{\nu \rho} \big) \dfrac{u^\mu}{\Theta}\\
& + \alpha_0 \Pi \dfrac{q^\mu}{\Theta} + \alpha_1 \dfrac{\Pi^{\mu \nu}q_\nu}{\Theta}.
\end{split}
\end{equation}
The quantity $s$ is the equilibrium entropy, i.e. the entropy which is computed using the equilibrium equation of state
\begin{equation}
s =s(\rho,n)
\end{equation}
of the fluid. This model is said to be a second-order theory because the entropy four-current is expanded to the second order in the deviations from local thermodynamic equilibrium. The second-order terms in $q^\mu,\Pi,\Pi^{\mu \nu}$ appearing in equation \eqref{secondoOrdine} are, indeed, the novelty introduced in the model and their presence can compensate (if the coefficients satisfy appropriate conditions) the inevitable explosion of the zeroth and first order contributions we described in the previous sections.

The equations of motions, which we do not need to report here, are constructed in way to ensure that
\begin{equation}\label{evrftjy}
\nabla_\mu s^\mu = \dfrac{q^\mu q_\mu}{k\Theta^2} + \dfrac{\Pi^2}{\zeta \Theta} + \dfrac{\Pi^{\mu \nu}\Pi_{\mu \nu}}{2\eta \Theta} \geq 0,
\end{equation}
where $k$, $\zeta$ and $\eta$ are respectively the heat conductivity, the bulk viscosity and the shear viscosity coefficient.

\subsection{The stability conditions}\label{BIS2}

Let us summarize the steps followed by \citet{Hishcock1983} to find the conditions for the second-order theory to admit a stable equilibrium.

First of all, they consider the stationary solutions to the system of equations, which are good candidates to be also equilibrium configurations. 
Clearly, for the configuration to be stationary, one needs to set the entropy production to zero. Thus from \eqref{evrftjy} they find that all the dissipative terms must vanish:
\begin{equation}\label{qawsdef}
q^\mu = \Pi = \Pi^{\mu \nu}=0.
\end{equation}
Therefore, the energy-momentum tensor of the fluid in equilibrium must be that of a perfect fluid, in agreement with kinetic theory \citep{cercignani_book,degroot_book}. 
They also use the equations of motion explicitly to obtain the remaining equilibrium conditions which are provided by kinetic theory: 
\begin{equation}
\nabla_\mu (u_\nu/\Theta)+\nabla_\nu (u_\mu/\Theta) =0 
\end{equation}
and
\begin{equation}
\nabla_\mu \bigg( \dfrac{\rho +P -\Theta s}{n\Theta} \bigg)=0.
\end{equation}
Once the candidates to be equilibrium states are found, they write the equations of motion for perturbations about these states. Subsequently, they show that it is possible to define a current $E^\mu$ (equation 44 in \cite{Hishcock1983}) which is quadratic in the deviations and  satisfies the condition
\begin{equation}\label{dEEE}
\nabla_\mu E^\mu = -\bigg[\dfrac{\delta q^\mu \delta q_\mu}{k\Theta^2} + \dfrac{(\delta \Pi)^2}{\zeta \Theta} + \dfrac{\delta\Pi^{\mu \nu}\delta\Pi_{\mu \nu}}{2\eta \Theta}   \bigg] \leq 0,
\end{equation}
where $\delta f$ is the perturbation of a generic hydrodynamic variable $f$. Introducing, then, the functional
\begin{equation}\label{EEEEEEt}
E = \int E^0 \, d_3 x,
\end{equation}
%If, now, one introduces an arbitrary time coordinate $t$, this defines a foliation of the spacetime into space-like 3D hypersurfaces $\Sigma(t)$ defined through the conditions $t=const$. Therefore one can introduce the functional
%\begin{equation}\label{EEEEEEt}
%E(t) = -\int_{\Sigma(t)} E^\mu \, d\Sigma_\mu
%\end{equation}
equation \eqref{dEEE} implies
\begin{equation}\label{hol1}
\dfrac{dE}{dt} \leq 0.
\end{equation}
Finally, they find the conditions under which 
\begin{equation}\label{hol2}
E \geq 0
\end{equation}
for any small deviation from equilibrium, obtaining a set of constraints for the equation of state and the coefficients of the model. When these constraints are satisfied, then, combining \eqref{hol1} and \eqref{hol2}, one obtains
\begin{equation}
E(t) \in [0,E(0)]  \spc \forall t \geq 0,
\end{equation}
implying, for arguments analogous to those we exposed by us in section \ref{SonoLya}, the Lyapunov stability of the configuration.

\subsection{Equivalence with the maximum entropy principle}

At this point, showing that the stability conditions found by \citet{Hishcock1983} imply that the entropy is maximum at equilibrium is straightforward.

First of all, we note that \eqref{qawsdef} implies
\begin{equation}
\delta q^\mu = q^\mu \spc \delta \Pi = \Pi  \spc \delta  \Pi^{\mu \nu} = \Pi^{\mu \nu}.
\end{equation}
Therefore we can combine \eqref{evrftjy} with \eqref{dEEE} to obtain
\begin{equation}\label{identico!}
\nabla_\mu s^\mu = -\nabla_\mu E^\mu.  
\end{equation}
Considering that $S$ and $E$ are given respectively by \eqref{bucatini} 
%\begin{equation}\label{pappagallo2}
%S(t) = -\int_{\Sigma(t)} s^\mu d\Sigma_\mu
%\end{equation}
%with the total entropy of the system. Therefore 
and \eqref{EEEEEEt}, equation \eqref{identico!} implies
\begin{equation}\label{chiappa}
S(t)+E(t) = S(0)+E(0).
\end{equation}
Now we impose that the theory is stable and that all the perturbations (which conserve the original value of the constants of motion of the fluid) decay to zero as $t \longrightarrow +\infty$, implying
\begin{equation}
S(+\infty)=S_{\text{eq}}  \spc E(+\infty)=0.
\end{equation}
The second condition results from the fact that $E^\mu$ is quadratic in the perturbation and therefore is exactly zero in equilibrium. The quantity $S_{\text{eq}}$ is simply the value of the entropy in the unperturbed equilibrium state. If we plug these two conditions in \eqref{chiappa}, we obtain
\begin{equation}
S_{\text{eq}} =S(0)+E(0),
\end{equation}
which plugged again into \eqref{chiappa} implies
\begin{equation}
E = S_{\text{eq}} -S.
\end{equation}
Since this relation holds for any initial (dynamically accessible) small perturbation, we have proved that the functional $E$ is nothing but the second-order correction to the entropy in the deviations from equilibrium. The stability condition \eqref{hol2}, then, implies
\begin{equation}
S_{\text{eq}} \geq S,
\end{equation}
which is the requirement that the entropy is maximum in the equilibrium state and is, therefore, a Lyapunov function.

\subsection{Example: stable heat conduction}

It is interesting to verify how the second-order terms in the entropy current counterbalance the explosion of the first-order theories with a concrete example. We consider again the Eckart fluid  introduced in subsection \ref{the_prrof}, but we replace the entropy current \eqref{entropppP} with the Israel-Stewart prescription
\begin{equation}\label{zszswew}
s^\mu = su^\mu + \dfrac{1}{\Theta} q^\mu - \dfrac{\beta_1 q^\nu q_\nu}{2\Theta}  u^\mu .
\end{equation}
The stability condition (i.e. the condition to have $E \geq 0$) obtained by \cite{Hishcock1983} for this model reads
\begin{equation}\label{BeTTa1}
\beta_1 > \dfrac{1}{\rho +P}.
\end{equation}
Let us verify that this makes the state $v=0$ (i.e. the homogeneous perfect fluid configuration) a local maximum of the entropy, as discussed in section \ref{the_prrof}.

Since the stress-energy tensor is the same, the steps which lead us to equation \eqref{S0} are unchanged, apart from the fact that we need to add to $s^0$ the second order contribution, obtaining
\begin{equation}\label{ninetifive}
s^0 = \tilde{s} \, \bigg( 1+ \dfrac{v^2}{2} \bigg)- \dfrac{\beta_1 q^1 q^1}{2\Theta}  .
\end{equation}
We have made the replacement $q^\nu q_\nu = q^1 q^1$ because $-q^0 q^0$ is a fourth order term, see equation \eqref{q0}. The condition that the entropy is maximum in equilibrium reads
\begin{equation}
\tilde{s} \geq s^0,
\end{equation}
which, with the aid of \eqref{sbup}, implies
\begin{equation}\label{beTa1}
\beta_1 > \tilde{s}\Theta \bigg(\dfrac{4}{3} \mathcal{E} \bigg)^{-2}.
\end{equation}
All the quantities appearing in the inequality above are background terms (as can be seen from \eqref{ninetifive}), thus they can be evaluated at $v=0$, which, from \eqref{equillll}, we know to be characterized by the conditions 
\begin{equation}
\rho = \mathcal{E} \spc s= \tilde{s} .
\end{equation}
However, from \eqref{kin} and \eqref{UEU}, we also have the chain of identities
\begin{equation}
s\Theta = \dfrac{4}{3} \rho = \rho +P.
\end{equation}
With these conditions it is possible to prove the equivalence of \eqref{beTa1} with \eqref{BeTTa1}. We have thus verified that the condition of dynamic stability of the Israel-Stewart theory coincides with the condition of maximum entropy in equilibrium.

\subsection{The problem of the instability for large deviations}

%Our thermodynamic study allows us also to understand the origin of the instability of Israel-Stewart for large deviations from equilibrium. 

Equation \eqref{BeTTa1} is the condition for $\tilde{s}$ to be the maximum value of $s^0$ for small deviations from equilibrium. However, it is not guaranteed that this maximum is absolute. In fact, \citet{Hishcock1988} have shown that, for sufficiently large deviations from equilibrium, the system can still admit runaway solutions. 
Let us analyse the behaviour of $s^0$, given in \eqref{zszswew}, for an arbitrarily large $v$.

\begin{figure}
\begin{center}
	\includegraphics[width=0.5\textwidth]{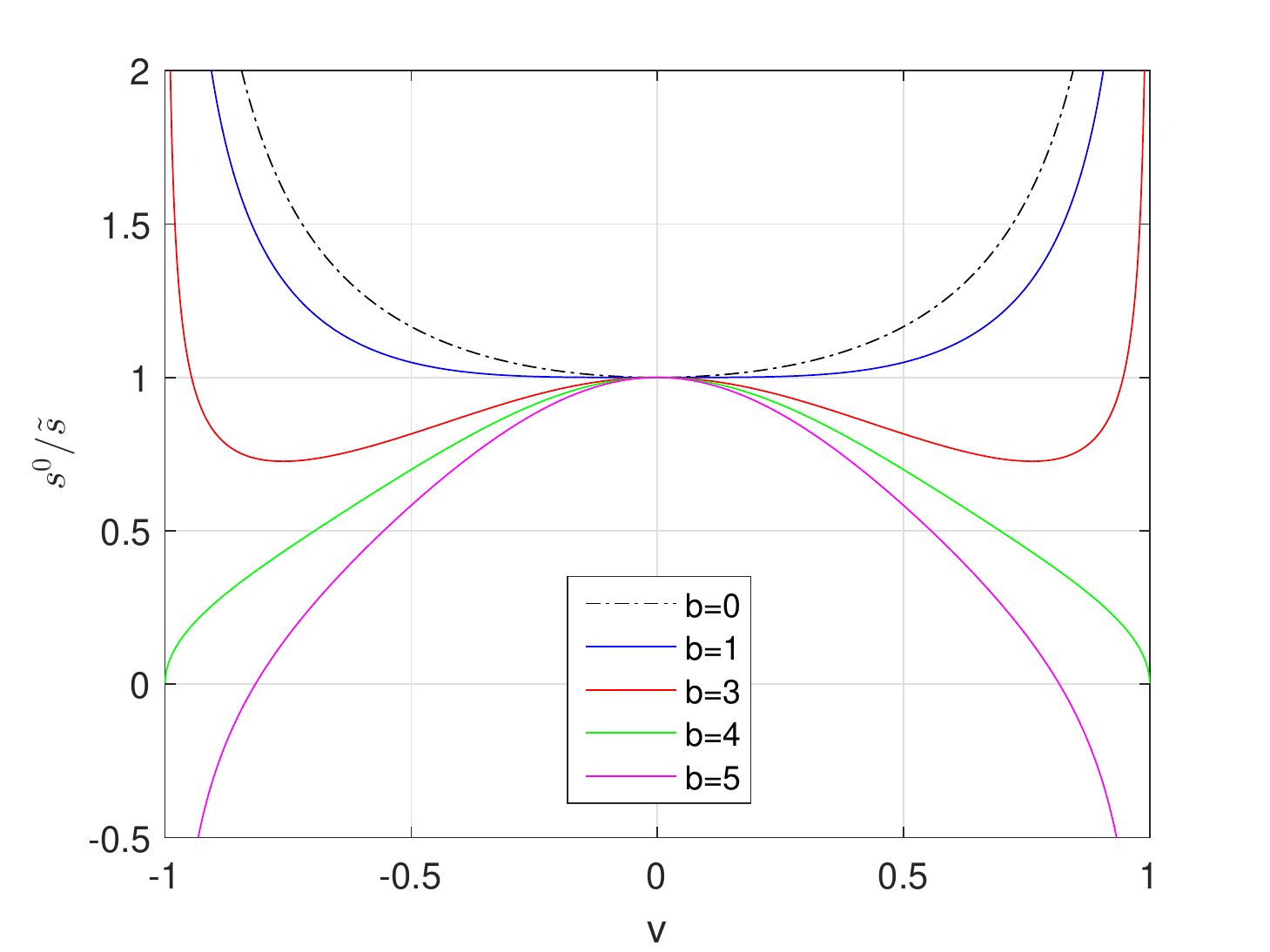}
	\caption{
	The normalised Israel-Stewart entropy per unit volume $s^0/\tilde{s}$ as a function of the velocity $v$, as given by equation \eqref{iltutto}, for $b=0,1,3,4,5$. We see that the local stability condition $b>1$ obtained by \citet{Hishcock1983} corresponds to the requirement that the equilibrium state $v=0$ is a local maximum of the entropy. However, if we want to ensure the global stability, this maximum must be a global maximum and we need to impose the stronger requirement $b \geq 4$.}
	\label{fig:second}
	\end{center}
\end{figure}

We can split the entropy density as 
\begin{equation}\label{In2duw}
s^0 = s^0_I + s^0_{II},
\end{equation}
where $s^0_I$ is the zeroth+first order contribution, given by equation \eqref{gringo}, and 
\begin{equation}\label{soii}
s^0_{II} = -\dfrac{\beta_1 \gamma}{2\Theta} (q^1 q^1-q^0 q^0)
\end{equation} 
is the second-order contribution. To make a parametric study, we take as a prescription for $\beta_1$ the generic expression
\begin{equation}
\beta_1 = \dfrac{b}{\rho +P},
\end{equation}
where $b \geq 0$ is a free constant coefficient. With calculations which are analogous to those made in section \ref{EcKo} one can obtain the expression
\begin{equation}
\dfrac{s^0_{II} }{\tilde{s}} = - \dfrac{b }{2} v^2 \bigg( 1-v^2 \bigg)^{-1/2} \bigg( 1+v^2 \bigg)^{-5/4} \bigg(1-\dfrac{v^2}{3} \bigg)^{-3/4}. 
\end{equation} 
Plugging this formula into \eqref{In2duw} and recalling \eqref{gringo}, we obtain
\begin{equation}\label{iltutto}
\dfrac{s^0}{\tilde{s}} =\bigg( 1+\dfrac{2-b}{2}v^2 \bigg) \bigg( 1-v^2 \bigg)^{-1/2} \bigg( 1+v^2 \bigg)^{-5/4} \bigg(1-\dfrac{v^2}{3} \bigg)^{-3/4}. 
\end{equation}
In figure \ref{fig:second} we show how the profile of $s^0/\tilde{s}$ varies for different values of $b$. 
For $b<1$ the stability condition \eqref{BeTTa1} for small deviations from equilibrium is not fulfilled and the state $v=0$ is a minimum. The situation $b=1$ is the case in which the second order expansion of $s^0$ around $v=0$ is zero. We see that the next non-vanishing order in $v$, the fourth order, is still positive: the perfect fluid state is still a minimum, and therefore is unstable. 
For $1<b<4$ the second order term of the expansion is negative, thus $v=0$ is a maximum. However, it is a local maximum, because for large $v^2$ the positive divergence of $s_I^0$ still dominates over the negative contribution of $s_{II}^0$. 
For the critical value $b=4$ the two divergences compensate each other exactly and we have $s^0 \rightarrow 0$ as $v\rightarrow \pm 1$. 
For $b \geq 4$ the theory is stable for any homogeneous deviation from equilibrium and $v=0$ is likely to be  the absolute maximum of the entropy (to be certain one should also make a study involving non-homogeneous configurations).

Therefore, we have shown that the condition \eqref{BeTTa1} is not enough to guarantee that the entropy is a Lyapunov function over the whole configuration space, but there is an interval of values of the parameter $b$ ($1<b<4$) in which the homogeneous perfect fluid is only a point of local maximum.

Again, we remark that to study the stability of the theory there is no need to know the details of the equations of motion of the fluid. In fact, we have not invoked their hyperbolic nature or their telegraph-type form. The only property needed is to know that they are constructed in a way to ensure that the entropy current \eqref{zszswew} has a non-negative divergence.

\subsection{The effect of the particle conservation}\label{sisisisis}

In their analysis, \citet{Hishcock1988}  consider the case $b=5$ and, even if they are working with an ultra-relativistic gas and they impose $b>4$, they still find that the model is unstable for large deviations from equilibrium. The key difference with our case is that they have a conserved particle number, while we are working at zero chemical potential. The conservation of the particle number has the effect of increasing the instability for large $v$. 

To show this, we first note that \eqref{conservo} implies in the homogeneous case
\begin{equation}
n^0 = \gamma  n = \text{const}.
\end{equation}
Now, since equation \eqref{system2} is a consequence only of the structure of the energy-momentum tensor and not of the equation of state, it holds also when the particle number is conserved. Therefore, assuming the ideal gas law (valid in the non-degenerate limit)
\begin{equation}\label{IIIIIdeale}
P = n\Theta,
\end{equation} 
we obtain
\begin{equation}\label{instabo}
\Theta =\gamma \, \dfrac{\mathcal{E} }{n^0} \, \dfrac{1+v^2}{3-v^2},
\end{equation}
which is equation (15) of \cite{Hishcock1988}. This relation implies that the temperature diverges when $v^2 \longrightarrow 1$. This is the key difference with respect to our case at zero chemical potential, because in that case all the rest-frame thermodynamic quantities remain finite also when the fluid approaches the speed of light. 

To see the implications on the stability, we insert \eqref{instabo} into \eqref{soii}. As a result, the Lorentz factors cancel out and $s_{II}^0$ is now finite even for $v^2 = 1^-$. The second-order contribution to the entropy is, therefore, no longer able to compensate the divergence of the first-order term $s_I^0$ and the fluid is unstable for any $b$ when sufficiently large deviations from equilibrium are considered, in agreement with \cite{Hishcock1988}. 

\begin{figure}
\begin{center}
	\includegraphics[width=0.5\textwidth]{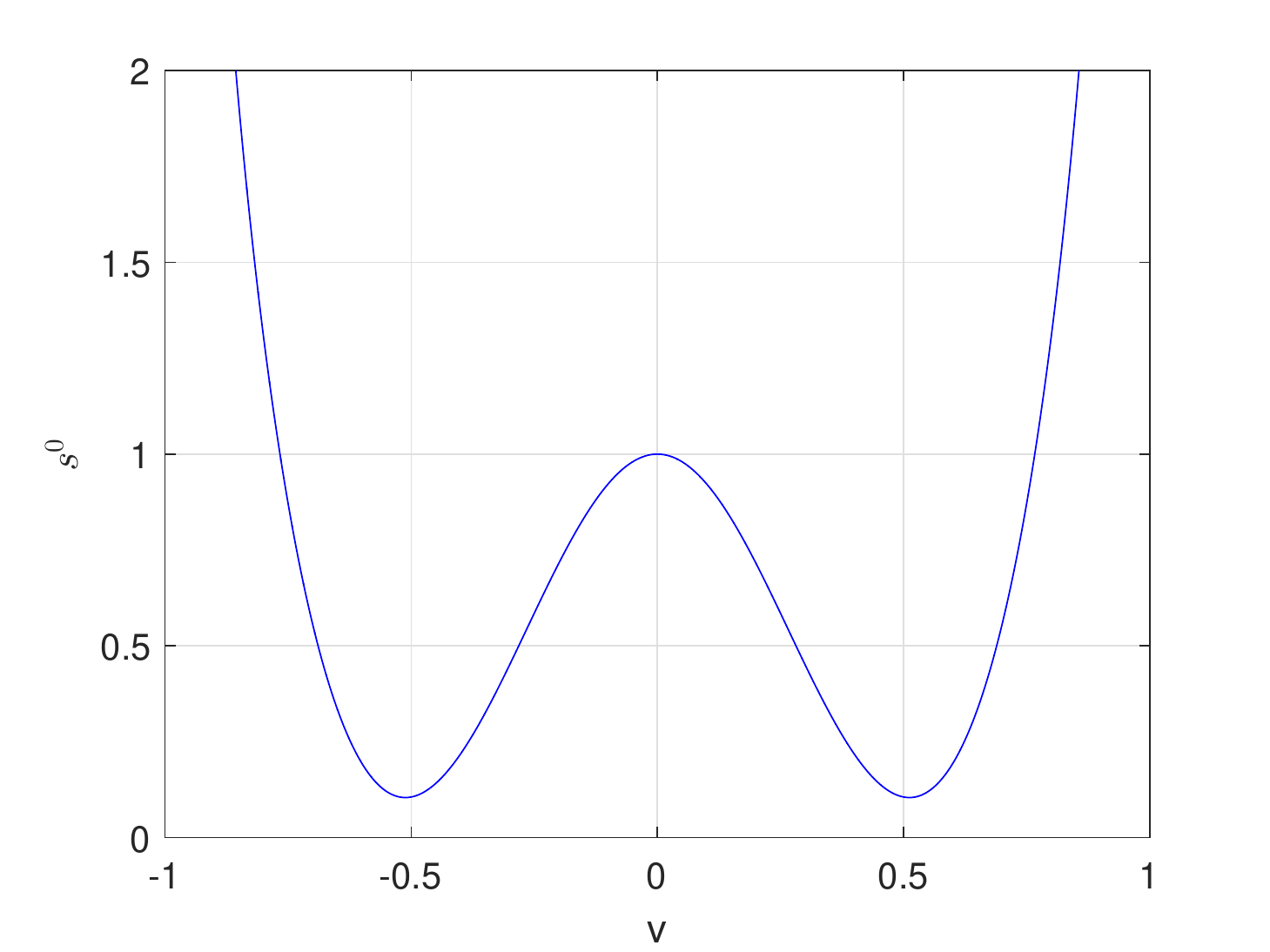}
	\caption{Israel-Stewart entropy density $s^0$, given in equation \eqref{zszswew}, for a non-degenerate ultra-relativistic homogeneous heat-conducting gas with conserved particle number, zero total momentum and $\beta_1 = 5/(\rho+P)$. This is the case considered by \citet{Hishcock1988}. We have imposed $\mathcal{E}=n^0=1$ and we have set the units in a way that $s^0(0)=1$. As can be seen, $v=0$ is only a local maximum of the entropy density and this is the origin of the instability for large deviations from equilibrium.}
	\label{fig:Hish}
	\end{center}
\end{figure}

To complete the comparison, in figure \ref{fig:Hish} we show the profile of $s^0(v)$ for the fluid considered by \cite{Hishcock1988}. The analytical formula is not reported here, but it can be easily obtained by following analogous steps to those which lead to \eqref{iltutto}, knowing that the equilibrium entropy per particle associated with the equation of state \eqref{IIIIIdeale} is 
\begin{equation}
\dfrac{s}{n} = - \ln \bigg( \dfrac{n}{\Theta^3} \bigg) + \text{const}.
\end{equation}
We see from figure \ref{fig:Hish} that $s^0(v)$ has a minimum in $|v|=v_c:=0.51188$, which is the critical velocity at which the fluid becomes unstable. Now the thermodynamic origin of the instability of Israel-Stewart is clear: for $|v|<v_c$ the direction of positive entropy growth points towards $v=0$, thus $|v|$ decreases. On the other hand, for $v>v_c$, the speed has to increase to enforce the validity of the second law. Thus, the bifurcation found by \cite{Hishcock1988} at $|v|=v_c$ is a direct result of the entropy profile given in figure \ref{fig:Hish} and of the strict obedience of the system to the second law of thermodynamics. 

\section{First-order theories in more general frames}
\label{Norio}

There is increasing attention on new first-order theories which have been proved to be causal and stable, if the transport coefficients are appropriately tuned \citep{Bemfica_2019_nonlinear,Kovtun2019}. The idea is to use alternative hydrodynamic frames to both Eckart's and Landau-Lifschitz \citep{Bemfica_2018_conformi,Hoult2020}. In this final section we show that the stability of these theories is enforced by allowing for small violations of the the second law of thermodynamics. 

\subsection{A model for bulk viscosity}

To simplify our analysis we restrict our attention to a purely bulk-viscous fluid. The stress-energy tensor we consider is
\begin{equation}
T^{\mu \nu} = (\rho +P +\mathcal{A})u^\mu u^\nu + P g^{\mu \nu}.
\end{equation}
The equilibrium perfect-fluid energy-momentum tensor is still the one we presented in section \ref{genericoAed}. The deviations from equilibrium, instead of being located in a viscous stress contribution, are given by a correction $\mathcal{A}$ to the internal energy of the fluid,
\begin{equation}
\mathfrak{T}^{\mu \nu} = \mathcal{A} \, u^\mu u^\nu.
\end{equation}
Instead of the standard dependence of the viscous stresses on $\nabla_\mu u^\mu$ only, the non-equilibrium correction to the energy density is assumed to be 
\begin{equation}\label{brakiut}
\mathcal{A} = \chi_1 \dfrac{u^\mu \nabla_\mu \rho}{\rho + P} +\chi_2 \nabla_\mu u^\mu.
\end{equation}
The transport coefficients $\chi_1$ and $\chi_2$ are functions of $\rho$. 
The entropy current is assumed to have the form
\begin{equation}\label{jjjj}
s^\mu = \bigg( s+ \dfrac{\mathcal{A}}{\Theta}  \bigg) u^\mu,
\end{equation}
which, then, implies that the non-equilibrium correction is
\begin{equation}
\sigma^\mu =  \dfrac{\mathcal{A}}{\Theta}  u^\mu.
\end{equation}
By using the conservation of the energy-momentum ($\nabla_\mu T^{\mu \nu}=0$), it can be shown that the equation for the entropy production reads
\begin{equation}\label{procucimale}
\Theta \nabla_\mu s^\mu = -\mathcal{A} \dfrac{u^\mu \nabla_\mu \Theta}{\Theta} .
\end{equation}
It is now immediate to see that to ensure the validity of the second law \eqref{bnigckfml} for any fluid configuration we need to impose
\begin{equation}\label{nonloro}
\chi_1 \leq 0  \spc \chi_2 =0. 
\end{equation}
However, following the argument of \citet{Kovtun2019}, one may consider that, since \eqref{jjjj} is just an approximate formula for the entropy current, its four-divergence needs to be only approximately non-negative. In particular, if one considers that the deviations from equilibrium are small, then the effect of the dissipative terms on the hydrodynamic evolution are small and we can impose the perfect-fluid relation
\begin{equation}\label{bhimogvrnefvo}
u^\mu \nabla_\mu \rho \approx -(\rho+P)\nabla_\mu u^\mu
\end{equation}
as approximately satisfied. Therefore, using the definition \eqref{brakiut}, at the lowest order we can replace $\mathcal{A}$ in \eqref{procucimale} with the approximate expression   
\begin{equation}\label{mleko}
\mathcal{A} \approx (\chi_1-\chi_2) \dfrac{u^\mu \nabla_\mu \rho}{\rho + P}.
\end{equation} 
The constraint of non-negative entropy production, then, requires only
\begin{equation}\label{nhjmkilo}
\chi_1 -\chi_2 \leq 0.
\end{equation}
As can be seen, this is a much weaker constraint than \eqref{nonloro}. Indeed, the only way to make the theory stable is to impose that the two separate conditions \eqref{nonloro} are both violated, while keeping \eqref{nhjmkilo} fulfilled. 
In particular, \citet{Bemfica_2019_nonlinear} have derived the stability conditions  
\begin{equation}\label{maseiserio?}
\chi_2 > \chi_1 >0.
\end{equation}
Thus we see that a necessary condition of stability is that the second law is not exactly respected, but it holds only as an approximation (see also \cite{Poovuttikul2019}). Our goal in the next subsection is to explain why this is the case.

\subsection{Thermodynamic analysis}

\begin{figure}
\begin{center}
	\includegraphics[width=0.5\textwidth]{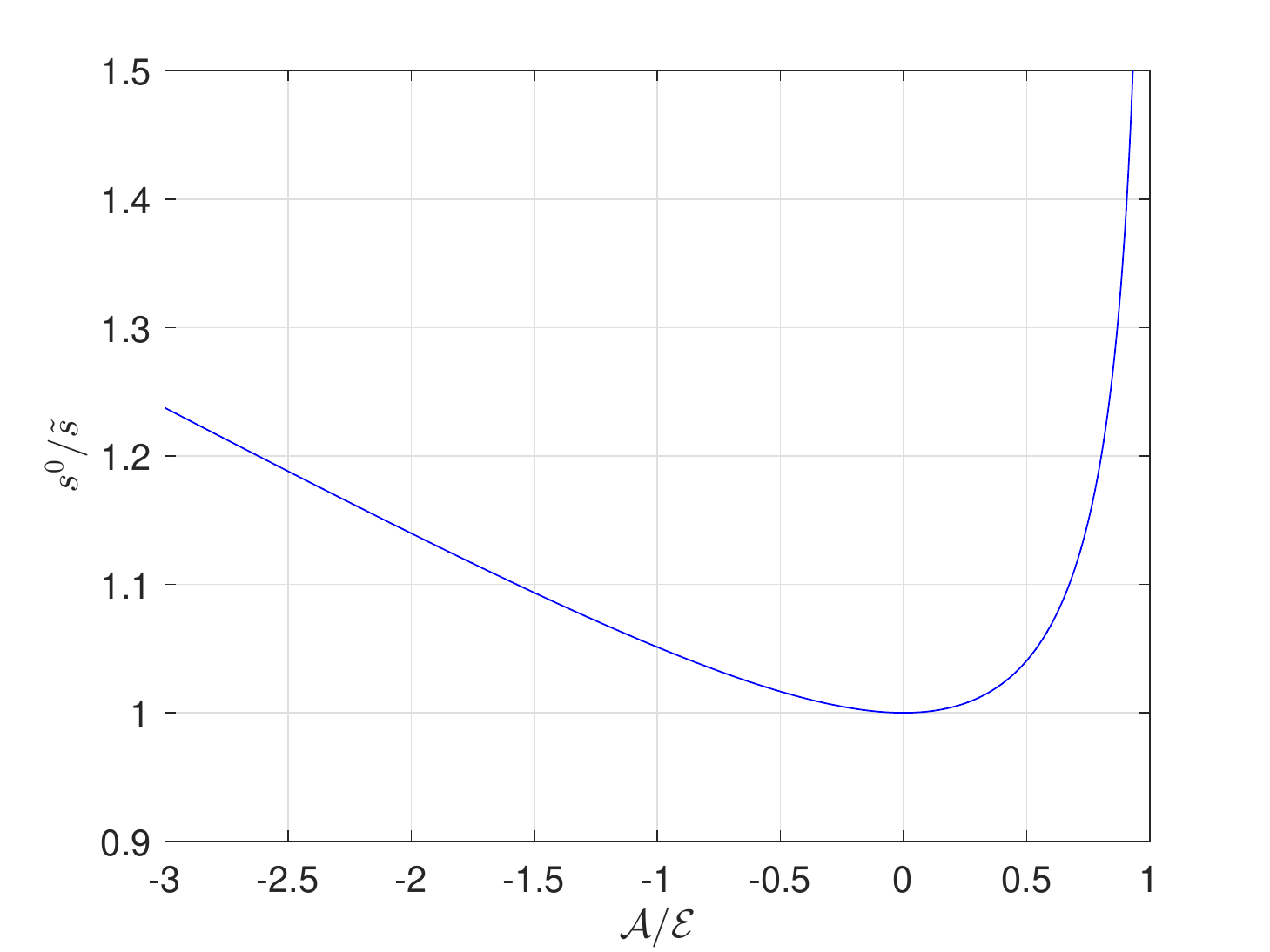}
	\caption{
	The normalised entropy per unit volume $s^0/\tilde{s}$ as a function of the dimensionless parameter $\mathcal{A}/\mathcal{E}$ as given by equation \eqref{sblearf}. 
	For $\mathcal{A}=0$ the fluid is in local thermodynamic equilibrium. We see that, also in this case, the entropy, restricted to the manifold of the dynamically allowed states, is not maximal in equilibrium. Thus if the second law was exactly respected, the theory of \citet{Bemfica_2019_nonlinear} would necessarily be unstable.}
	\label{fig:noro}
	\end{center}
\end{figure}

Similarly to what we did in the previous sections, we consider a homogeneous portion of the fluid and we impose for simplicity $\mathcal{P}^j=0$. Then it is immediate to see that there is no way for the fluid to change its velocity.  This implies that the relevant degrees of freedom of the fluid element are only two ($\rho$ and $\mathcal{A}$) and we have only one relevant constraint $(\mathcal{E})$. Again, we are dealing with a 1D manifold of dynamically allowed states. We parametrize it with $\mathcal{A}$ and we use the constraint equation \eqref{CCCooonnn} to write $\rho$ as a function of $\mathcal{A}$:
\begin{equation}\label{envokc}
\rho = \mathcal{E}-\mathcal{A}.
\end{equation}
From \eqref{jjjj}, we immediately obtain
\begin{equation}
s^0 = s + \dfrac{\mathcal{A}}{\Theta}.
\end{equation}
With calculations analogous to those made in the previous sections we can obtain the formula for the entropy density as a function of the parameter of the curve:
\begin{equation}\label{sblearf}
s^0 = \tilde{s} \, \bigg( 1-\dfrac{\mathcal{A}}{4\mathcal{E}} \bigg) 
\bigg( 1-\dfrac{\mathcal{A}}{\mathcal{E}} \bigg)^{-1/4} . 
\end{equation}
The graph of this function is shown in figure \ref{fig:noro}.

Also in this case the perfect fluid state (given by $\mathcal{A}=0$) is a saddle point of the entropy and not its maximum. Therefore, if we enforce the entropy production to be strictly non-negative, a perturbation which makes the system start with $\mathcal{A}\neq 0$ is not allowed to decay but can only grow. This tells us that it is necessary to violate the second law of thermodynamics to ensure that the perfect fluid state is a stable equilibrium configuration.

Let us study more in detail the evolution of small homogeneous perturbations around $\mathcal{A}=0$. 
We consider perturbations of the generic hydrodynamic function $f$ to have the form 
\begin{equation}
\delta f \, e^{\Gamma t},
\end{equation}
where $\delta f$ is an infinitesimal constant, and we impose that these perturbations preserve the constraints (thus they are dynamically allowed). From \eqref{brakiut} we can write the perturbation $\mathcal{A}$ at the first order as
\begin{equation}
\delta \mathcal{A} = \dfrac{3\chi_1}{4\mathcal{E}} \Gamma \delta \rho.
\end{equation} 
Note that the perturbation of $\nabla_\mu u^\mu$ is exactly zero because the fluid remains at rest. If we vary \eqref{envokc} we obtain $\delta \rho = -\delta \mathcal{A}$, which, plugged in the equation above, gives
\begin{equation}
\Gamma = - \dfrac{4\mathcal{E}}{3\chi_1}.
\end{equation}
The condition for the perturbation to decay is $\Gamma<0$, which implies $\chi_1 >0$. Thus we recover one of the two stability conditions \eqref{maseiserio?} found by \citet{Bemfica_2019_nonlinear}. On the other hand, if we want the second law to be strictly satisfied, then we need impose $\Gamma \geq 0$, which in turn implies $\chi_1 <0$. 
In this case we recover the first condition of \eqref{nonloro}.

The reason why in this configuration there is such a strong contradiction between the stability criteria and the second law is that we have constructed the system in a way to enforce $\nabla_\mu u^\mu =0$, while imposing $u^\mu \nabla_\mu \rho \neq 0$. Thus, the approximate relation \eqref{bhimogvrnefvo} does not hold for this initial condition and \eqref{mleko} is no longer valid. 

We remark that the amount of entropy which is annihilated is a second order in $\mathcal{A}$. However, the entropy current \eqref{jjjj} is only an approximation to the first order in $\mathcal{A}$ of the physical entropy current. Then, as we explained in section \ref{HowToslove}, this violation should not be considered a reason of concern, but, instead, represents the fundamental origin of the success of the theory.

\section{Conclusions}

In relativity, spatial gradients become linear combinations of derivatives in both space and time when one moves from one reference frame to the other. The number of degrees of freedom of the relativistic models of dissipation is, then, larger with respect to the Newtonian counterpart. This implies that the state-space has a bigger dimension and hosts a wide set of configurations which do not have a Newtonian analogue. We have shown that the instability of the \citet{Eckart1940} and \citet{landau6} theories arises from the fact that it is always possible to find a dynamically allowed path in this extended state space along which the total entropy grows with no bound. 

This fact tells us that the instability of both Eckart and Landau-Lifshitz theories has a purely thermodynamic nature: the assumed equilibrium state of the fluid (such as, for example, an homogeneous perfect fluid configuration) is not the maximum of the total entropy, but only a saddle point. The unstable modes evolve along the directions in the state space in which the entropy takes larger values with respect to the value assumed in the (supposed) equilibrium state. 
This also shows that the reason why these fluids do not obey the Onsager regression hypothesis (as has been pointed out by \citet{Garcia2009}) is that an equilibrium state does not exist and the entropy is not a Lyapunov function of the system.

We have then proved that the conditions for stability obtained by \citet{Hishcock1983} are the requirements for the second-order theory to produce a system with a maximum entropy state (for small deviation from equilibrium). The second-order contributions act in a way to compensate the explosion of the fist-order theory. In addition, we have seen that the same study can also be used to prove the possible instability of the \citet{Israel_Stewart_1979} theory for large deviations from equilibrium. In fact, we have verified that the critical speed $v_c= 0.51188$ found by \citet{Hishcock1988} at which a homogeneous heat-conducting ultra-relativistic fluid becomes unstable marks a saddle point of the entropy above which the latter starts growing with no bound, producing the instability. Finally, we have shown that the first-order theory of \citet{Bemfica_2019_nonlinear} does not restore the concavity of the entropy, but deals with the stability problem by allowing for small violations of the second law. In this kind of first-order theories the strict obedience to the second-principle would lead again to the (non-physical) explosion of small perturbations.

The main message of the present analysis is that, assuming a first-order expansion of the entropy current, one is selectively removing second-order contributions to the total entropy of the fluid. In this way, the concavity properties of the entropy are altered and, as a consequence, the absolute maximum is typically converted into a (non-physical) saddle point. The instability, then, arises when the second law is imposed, enforcing the growth of the approximated entropy at all the orders. In this way, the thermodynamic principle which originally was ensuring the Lyapunov stability of the system is converted into the main source of instability, pushing the system along the non-physical growing branches which depart from the equilibrium state. 

There are only two possible solutions to this problem. The first is to retain all the second-order contributions (which leads to the higher order formulations like \citet{Israel_Stewart_1979} and \citet{carter1991}) the second is to break the second law of thermodynamics at the second order (which leads to the first-order theories in more general frames proposed by \citet{Bemfica_2019_nonlinear}).   

Finally, our discussion also clarifies that in relativity the approach of extended irreversible thermodynamics \citep{Stewart_1977,Jou_Extended}, which promotes the dissipative terms to degrees of freedom, is a mathematical necessity. Even the first-order theories, which are not explicitly designed according to this philosophy, implicitly contain this assumption, which manifests itself as soon as the fluid is set into motion. Once this general fact is accepted, all the interpretative difficulties disappear and thermodynamics rules once again.

\section*{Acknowledgements}

We acknowledge support from the Polish National Science Centre grants SONATA BIS 2015/18/E/ST9/00577 and OPUS 2019/33/B/ST9/00942. Partial support comes from PHAROS, COST Action CA16214.

\appendix

\section{INSTABILITY OF THE DIFFUSION EQUATION IN RELATIVITY}
\label{AAA}

In this appendix we study the instability of the diffusion equation in special relativity. Despite its simplicity, this example provides physical intuition of how, changing reference frame, one might produce unexpected instabilities. 
A detailed analysis of the mathematical aspects of the problem can be found in \citet{Kost2000}, who proved that all the problems concerning the diffusion equation in relativity arise from its ill-posedness in the boosted frame. They also showed that the instability mechanisms of \citet{landau6} are formally identical to those of the diffusion equation, ensuring the generality of the results of the present appendix.

\subsection{The role of the relativity of simultaneity}

Consider a one-dimensional medium whose temperature field $\Theta$ obeys the diffusion equation (in rest frame of the medium)
\begin{equation}
\label{diffusionequation}
\dfrac{\partial \Theta}{\partial t} = \mathcal{D} \dfrac{\partial^2 \Theta}{\partial x^2},
\end{equation}
where $\mathcal{D}$ is, for simplicity, a constant. We ignore for the moment the issues related with causality, accepting the idea that signal propagation in this medium can be super-luminal. We are interested in the evolution of $\Theta$, as seen by an observer who is moving with velocity $v\neq 0$ with respect to the medium. The associated Lorentz transformation is
\begin{equation}\label{transf}
t' = \gamma (t-vx)  \spc  x' = \gamma (x-vt),
\end{equation}
with
\begin{equation}
\gamma = \dfrac{1}{\sqrt{1-v^2}} \, ,
\end{equation}
implying
\begin{equation}\label{parziale}
\dfrac{\partial}{\partial t} = \gamma \bigg( \dfrac{\partial}{\partial t'} - v \dfrac{\partial}{\partial x'} \bigg) \spc \dfrac{\partial}{\partial x} = \gamma \bigg( \dfrac{\partial}{\partial x'} - v \dfrac{\partial}{\partial t'} \bigg) \, .
\end{equation}
In the boosted frame, equation \eqref{diffusionequation} now reads 
\begin{equation}\label{cambiareeeeee}
\dfrac{\partial \Theta}{\partial t'} -v \dfrac{\partial \Theta}{\partial x'} = \mathcal{D} \gamma \bigg( \dfrac{\partial^2 \Theta}{\partial {x'}^2} - 2v\dfrac{\partial^2 \Theta}{\partial t' \partial x'} + v^2 \dfrac{\partial^2 \Theta}{\partial {t'}^2}  \bigg).
\end{equation}
Note that  equation  \eqref{diffusionequation} was of the first order in the reference frame of the medium, but it becomes of the second order in any other reference frame, due to the presence of the third term in the right-hand side of \eqref{cambiareeeeee}. 
Therefore, if in the rest-frame of the medium the state is entirely specified once we know the value of $\Theta$ everywhere at the initial time, in any other frame we need to know both $\Theta$ and $\partial \Theta/\partial t'$. The principle of relativity has enlarged the state-space, increasing the number of degrees of freedom of the system, as we discussed in subsection \ref{genericoAed}.

The origin of the problem is the relativity of simultaneity \citep{special_in_gen}, namely the fact that events which are simultaneous in a given reference frame may not be simultaneous in an other one. 
In fact, if there was an absolute notion of simultaneity, which would imply that $t'=t'(t)$, we would have
\begin{equation}
\dfrac{\partial}{\partial x} = \dfrac{\partial x'}{\partial x} \bigg|_t \dfrac{\partial}{\partial x'} + \dfrac{\partial t'}{\partial x}  \bigg|_t \dfrac{\partial}{\partial t'} = \dfrac{\partial x'}{\partial x}  \bigg|_t \dfrac{\partial}{\partial x'}
\end{equation}
and the second derivative in time in equation \eqref{cambiareeeeee} would not appear.

In some astrophysical contexts \citep{FragileKlusniak2018}, diffusion-type models for relativistic dissipation are included in numerical simulations assuming that in the reference frame considered in the simulation (which does not necessarily coincide with the rest-frame of the medium) the evolution can be approximated as quasi-static. This assumption is then used to neglect the term $\partial_{t'}^2 \Theta$ in \eqref{cambiareeeeee}. The resulting system is, then, not structurally different from a Newtonian model (it has the same number of degrees of freedom) and stability can be restored.

\subsection{Stability analysis}

Let us study the evolution of homogeneous configurations in the frame which is moving with respect to the medium. Equation \eqref{cambiareeeeee} reduces to
\begin{equation}\label{adessoesplode}
\dfrac{\partial \Theta}{\partial t'}  =\mathcal{D} \gamma  v^2 \dfrac{\partial^2 \Theta}{\partial {t'}^2},
\end{equation}
whose general solution is
\begin{equation}
\Theta (t') = \Theta_0 + \dfrac{\dot{\Theta}_0}{\Gamma_+} \big(e^{\Gamma_+ t'}-1 \big) ,
\end{equation}
where we have defined
\begin{equation}
\Gamma_+ = \dfrac{1}{\mathcal{D} \gamma  v^2} >0.
\end{equation}
The space of the initial conditions is determined by two parameters ($\Theta_0$ and $\dot{\Theta}_0$), instead of only one. Furthermore, we see that, whenever $\dot{\Theta}_0 \neq 0$, the solution diverges for large times. Hence, we have verified that the existence of unstable solutions arises directly from the possibility of setting the time-derivative of $\Theta$ freely and is, therefore, a pure consequence of the relativistic extension of the state-space.

To understand how the instability can develop in a boosted reference frame, while it does not appear in the frame of the medium, let us examine solutions given by initial conditions of the type
\begin{equation}
\dot{\Theta}_0 = \Gamma_+ \Theta_0.
\end{equation} 
Going to the medium rest-frame, using the transformation rule \eqref{transf}, we immediately see that these solutions have the form
\begin{equation}
\Theta (t,x) = \Theta_0 e^{\Gamma_+ \gamma (t-vx)},
\end{equation}
which is obviously a solution of \eqref{diffusionequation}. We see that along surfaces at constant time the space dependence of $\Theta$ is 
\begin{equation}
\Theta (t,x) \propto e^{-\Gamma_+ \gamma v x}.
\end{equation}
This means that (assuming $v>0$ for definiteness) the unstable solutions in the boosted frame correspond to configurations in the frame of the medium in which an infinite amount of energy is shifting uniformly to the right in the spacetime diagram, coming from $x =-\infty $. This again shows how the relativity of simultaneity, making even the notion of homogeneity frame-dependent, is a key ingredient to make the instability possible.

It is interesting to note that in the Newtonian limit ($v \longrightarrow 0$) the growth rate diverges (and does not go to zero as one might intuitively think), $\Gamma_+ \longrightarrow +\infty$. 
This happens also in the theories of \cite{Eckart1940} and \cite{landau6}. The reason is that, since the Newtonian theory has less degrees of freedom than the relativistic one, we cannot obtain it just by taking the limit of the equations of motion, but we also need to make a particular choice for the initial conditions. 

This can be understood by considering again equation \eqref{adessoesplode}, evaluated at $t'=0$:
\begin{equation}
\dfrac{\partial^2 \Theta}{\partial {t'}^2} (0) = \dfrac{ \dot{\Theta}_0}{\mathcal{D} \gamma v^2}.
\end{equation}
If we take the limit $v \longrightarrow 0$, while keeping $ \dot{\Theta}_0$ fixed and finite, we find
\begin{equation}
\dfrac{\partial^2 \Theta}{\partial {t'}^2} (0) \longrightarrow \infty.
\end{equation}
We, thus, obtain a fast diverging solution. However, we know that in the Newtonian limit $ \dot{\Theta}_0$ cannot be set arbitrarily, but must be zero, as predicted by \eqref{diffusionequation}. Therefore, to obtain the Newtonian theory, we need to send $v \longrightarrow 0$, while selecting the initial condition for $\dot{\Theta}_0$ directly from the Newtonian equation \eqref{diffusionequation}. This gives the expected result:
\begin{equation}
\dfrac{\partial^2 \Theta}{\partial {t'}^2} (0) =0.
\end{equation}

\subsection{The Cattaneo hyperbolic model}

The Cattaneo equation \citep{cattaneo1958} is a  modified diffusion equation that ensures finite signal propagation speed. This equation, as shown by \citet{Israel_Stewart_1979}, arises naturally from a relativistic thermodynamic approach, and 
includes a relaxation term with a time-scale $\tau>0$, 
\begin{equation}\label{cattaneoequation}
\tau \dfrac{\partial^2 \Theta}{\partial t^2} +\dfrac{\partial \Theta}{\partial t} = \mathcal{D} \dfrac{\partial^2 \Theta}{\partial x^2}\, .
\end{equation}
It is known (see e.g. \cite{Kost2000}) that the foregoing equation (which now is of the second order in the time-derivative even in the rest-frame) admits a signal propagation which cannot exceed the speed
\begin{equation}
c_{II} = \sqrt{\dfrac{\mathcal{D}}{\tau}},
\end{equation} 
called second-sound speed. It is, then, clear that the theory is compatible with causality requirements if and only if
\begin{equation}\label{sublumo}
c_{II} \leq 1.
\end{equation}
We can easily verify that this modification solves also the stability problems. In fact, from \eqref{parziale}, we find
\begin{equation}
\tau \dfrac{\partial^2 \Theta}{\partial t^2} = \tau \gamma^2 \bigg( \dfrac{\partial^2 \Theta}{\partial {t'}^2} -2 v \dfrac{\partial^2 \Theta}{\partial t' \partial x'} + v^2 \dfrac{\partial^2 \Theta}{\partial {x'}^2} \bigg).
\end{equation}
The equation for the homogeneous solutions \eqref{adessoesplode}, then, becomes
\begin{equation}
\tau \gamma \dfrac{\partial^2 \Theta}{\partial {t'}^2} + \dfrac{\partial \Theta}{\partial t'}  =\mathcal{D} \gamma  v^2 \dfrac{\partial^2 \Theta}{\partial {t'}^2},
\end{equation}
whose general solution is
\begin{equation}
\Theta (t') = \Theta_0 + \dfrac{\dot{\Theta}_0}{\Gamma_-} \big(e^{\Gamma_- t'}-1 \big) ,
\end{equation}
with
\begin{equation}
\Gamma_- = \dfrac{1}{\gamma(\mathcal{D}v^2-\tau)}.
\end{equation}
The stability requirement is $\Gamma_- <0$, which (imposed for every $v^2<1$) implies \eqref{sublumo}.

\subsection{Connection between causality and stability}

\citet{Hishcock1983,Olson1990} have shown that, in Israel-Stewart-type theories, stability and causality are essentially equivalent. As we saw in the foregoing subsection, the Cattaneo equation is no exception and we can use it to give a simple physical intuition of this connection.

Let us consider a temperature profile of the form
\begin{equation}\label{rnviel}
\Theta = \Theta_0 e^{\Gamma\gamma (t-vx)} \spc \Theta_0>0.
\end{equation} 
Clearly, if $\Gamma>0$ the underlying theory of which this profile is solution is unstable in the boosted frame. 

Neglecting overall additive constants, the energy per unit length (measured in the rest-frame of the medium) is
\begin{equation}
\rho = c_v \Theta,
\end{equation}
where $c_v$ is the specific heat (note that this is one of the approximations one needs to invoke in order to derive \eqref{diffusionequation}). Now, let us define the function
\begin{equation}
\mathcal{E}(t) := \int_t ^{+\infty} \rho(x,t) dx,
\end{equation}
which measures the amount of energy contained in the half-line $x>t$ at the time $t$. Since the half-line $x>t_1$ is the causal past of $x>t_2$ (for $t_1 < t_2$), then, if the theory is causal, we need to have $\mathcal{E}(t_1) \geq \mathcal{E}(t_2)$, because no energy can be transferred to a region from outside its past light-cone. 
Therefore, if the theory is causal, it must be true that
\begin{equation}
\dfrac{d \mathcal{E}}{dt} \leq 0.
\end{equation} 
On the other hand, it is immediate to verify that
\begin{equation}
\mathcal{E}(t) = \mathcal{E}(0) e^{\Gamma \gamma(1-v) t},
\end{equation}
which proves that if the theory is unstable ($\Gamma>0$), then it is not causal ($d\mathcal{E}/dt>0$).

This argument shows that in a causal theory there is not enough energy to develop instabilities of the form \eqref{rnviel}. In fact, it is necessary to allow for superluminal transport of energy to transfer the energy from $x=-\infty$ to the bulk of the system sufficiently fast (in the rest-frame of the medium) to produce the instability.

%\bibliographystyle{mnras}
%\bibliography{Biblio}

\bibliography{Biblio}

\label{lastpage}

\end{document}